\documentclass{aa}
\usepackage{txfonts} 
\usepackage{graphicx}
\begin{document}
   \title{Search for nearby stars among proper motion stars
          selected by optical-to-infrared photometry}

   \subtitle{III. Spectroscopic distances of 322 NLTT stars\thanks{The full
version of Table~\ref{20obj} including all 322 objects and a corresponding file
with individual notes on the table entries are only available in electronic
form at the CDS via anonymous ftp to cdsarc.u-strasbg.fr (130.79.128.5) or
via http://cdsweb.u-strasbg.fr/cgi-bin/qcat?J/A+A/vol/page/}$^,$\thanks{Figures~\ref{bluestd} and \ref{redstd} are only available in electronic form at
http://www.edpsciences.org}}

   \author{R.-D. Scholz
          \inst{1,4}
          \and
          H. Meusinger
          \inst{2,4}
          \and
          H. Jahrei{\ss}
          \inst{3}
          }

   \offprints{R.-D. Scholz}

   \institute{Astrophysikalisches Institut Potsdam, An der Sternwarte 16,
              D--14482 Potsdam, Germany\\
              \email{rdscholz@aip.de}
         \and
             Th\"uringer Landessternwarte Tautenburg,
             D--07778 Tautenburg, Germany\\
             \email{meus@tls-tautenburg.de}
         \and
             Astronomisches Rechen-Institut, M\"onchhofstra{\ss}e 12-14,
             D--69120 Heidelberg, Germany\\
             \email{hartmut@ari.uni-heidelberg.de}
         \and
             Visiting astronomer, German-Spanish Astronomical Centre, 
             Calar Alto, operated by the Max-Planck-Institute for Astronomy, 
             Heidelberg, jointly with the Spanish National Commission for 
             Astronomy 
             }

   \date{Received ...; accepted ...}

   \abstract{Distance estimates based on low-resolution spectroscopy 
and 
Two Micron All Sky Survey (2MASS)
$J$ magnitudes are presented for a large sample of 
322
nearby candidates from Luyten's NLTT catalogue. Mainly relatively bright 
(typically $7 < K_\mathrm{s} < 11$) and red high proper motion stars  
have been
selected according to their 2MASS magnitudes and optical-to-infrared 
colours ($+1 < R-K_\mathrm{s} < +7$). Some LHS stars
previously lacking spectroscopy 
have also been
included. 
We have classified
the majority of the objects as early-M dwarfs (M2-M5). More than 70\%
of our targets turned out to lie within the 25~pc horizon of the 
catalogue of nearby stars, with 50 objects placed within 15~pc and
8 objects being closer than 10~pc. 
Three objects in the 10 pc sample have no previously published
spectral type: LP~876-10 (M4), LP~870-65 (M4.5), and LP~869-26 (M5).
A large fraction of the 
objects in our sample (57\%) 
have independent 
distance estimates, 
mainly by the recent efforts of Reid and collaborators. Our 
distance determinations are generally in good agreement with theirs.
11 rather distant ($d > 100$~pc) objects have also been 
identified, 
including
a probable halo, but relatively hot ($T_{\rm eff}\sim13000$~K) white dwarf 
(LHS~1200) and 10 red dwarfs with 
extremely large tangential velocities ($250 < v_\mathrm{t} < 1150$~km/s).
Altogether, there are 11 red dwarfs 
(including one within 70~pc) 
with tangential velocities larger
than about 250~km/s. All these objects are suspected to be
in fact subdwarfs, if so, their distances would be only about
half of our original estimates.
The  three most extreme objects in that respect 
are the K and early M dwarfs LP~323-168, LHS~5343 and LP~552-21
with corrected distances between 180~pc and 400~pc and resulting 
tangential velocities still larger than about 400~km/s. 
             
   \keywords{techniques: spectroscopic --
             surveys --
             astrometry --
             stars: distances --
             stars: late-type --
             Solar neighbourhood
               }
   }

\authorrunning{Scholz, Meusinger \& Jahrei{\ss}}
\titlerunning{Spectroscopic distances of 322 NLTT stars}
   \maketitle

\section{Introduction}

Whereas all Solar-type stars in the immediate neigbourhood of the Sun
($d<25$~pc)
are well known and accurately mapped after the successful Hipparcos mission
(ESA \cite{esa97}), our know\-ledge on the nearest lower-mass dwarf stars (and
probably on white dwarf stars, too) is by far not complete. 
Henry  et al.\ (\cite{henry02}) estimated the number of missing stellar systems
within 25~pc as high as 63\,\%. Within 10~pc the number of missing systems
is still likely about 25\,\% (Reid et al.\ \cite{reid03a}) or even more than
30\,\% (Henry  et al.\ \cite{henry02}).

Getting the nearby stellar sample completed is motivated by various
astronomical research areas. It has been realised for a long time that
the immediate Solar neighbourhood is the only region of a galaxy where there
is a real chance for a complete registration of its stellar content. A
complete
volume-limited stellar sample provides the groundwork for the knowledge of
the
local stellar luminosity function, the mass function of field stars, and
the Galactic star formation history (Reid, Gizis \& Hawley \cite{reid02c};
Gizis, Reid \& Hawley \cite{gizis02}; Cruz et al.\ \cite{cruz03}).
Moreover, the nearest representatives of a given class of stars can be
considered as benchmark sources allowing detailed follow-up studies.
In addition, a new interest in the local stellar population arises from
the field of extra-solar planet research.
Planet search programs based on Doppler spectroscopy first concentrated on
the
brighter F, G and K dwarfs. Late-type dwarfs
constitute only a small fraction in previous target samples
(e.g. Cochran et al. \cite{cochran02}; Endl et al. \cite{endl03}),
but there is an increasing interest in M dwarfs
(e.g.\, Butler et al.\ \cite{butler04}). Recently, it has been demonstrated
that Neptune-mass planets can be found around M dwarfs
(Marcy et al.\ \cite{marcy01}; Butler et al.\ \cite{butler04}).
M dwarfs are by far the most numerous Solar neighbours, and
the overwhelming majority of the missing nearby stars is also expected to
be of spectral types M.  Planets around nearby
low-mass dwarfs are good targets for accurate astrometric mass determinations
(Benedict et al. \cite{benedict02}), and
photometric transit searches for planets in habitable zones
are exceptionally sensitive for M dwarfs (Gould et al.
\cite{gould03b}). Finally, nearby, intrinsically faint stars will be prime
targets for the imaging of extrasolar planets by future
space-based missions. Direct imaging is a relatively "target
starved" field. Finding more nearby neighbours is actually very important
in this context.
   
Most of the stars within the 25~pc horizon of the
Catalogue of Nearby Stars, 3rd edition
(CNS3; Gliese \& Jahrei{\ss} \cite{gliese91}), were originally detected as
nearby candidates by their large proper motions. The Luyten Half Second (LHS) 
catalogue includes about 4500 stars with proper motions exceeding
0.5 arcsec/yr 
(Luyten \cite{luyten79})
. The New Luyten Two Tenths
(NLTT) catalogue of Luyten (\cite{luyten7980}) contains about 60000 stars
with proper motions larger than 0.18\,arcsec/yr. Due to the large
numbers of NLTT stars, only small sub-samples of the most remarkable objects 
with respect to different selection criteria (magnitudes, proper motions, 
colours) have been further investigated. 
Although the LHS stars 
have gotten
more attention,
there are still some LHS stars lacking spectroscopic classification.
Unfortunately, the original NLTT
colours are very crude estimates, and the positions were only given with very
low accuracy, which hampered the cross-identification with other catalogues
providing better photometry.

Special efforts were therefore needed in order to re-identify NLTT stars
and to select nearby candidates based on better colour measurements.
In two preceeding papers of this series, presenting the discovery of
three very  nearby ($d<10$~pc) objects 
(Scholz, Meusinger \& Jahrei{\ss}~\cite{scholz01}, hereafter paper~I, and
McCaughrean, Scholz \& Lodieu~\cite{mjm02}, hereafter paper~II), 
the combination of the NLTT with other optical
sky surveys and especially with the new infrared data from the
Two Micron All Sky Survey (2MASS; Cutri et al.\ \cite{cutri03})
was demonstrated to be an  effective tool for finding 
previously unknown nearby red dwarfs.

\begin{figure}[htb]
\begin{center}
\includegraphics[width=6.4cm, angle=270]{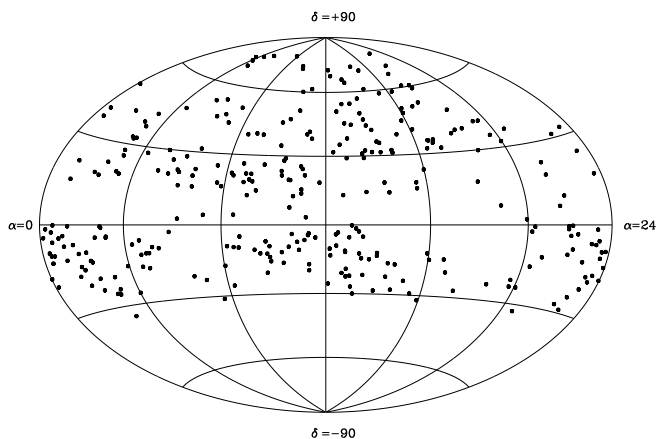}
\caption[Sky distribution]{Distribution in equatorial coordinates of the
322 NLTT stars with spectroscopic distance estimates obtained in this study.
}
\label{skydist}
\end{center}
\end{figure}

In the present paper we describe the results of spectroscopic 
follow-up observations
of a larger sample of NLTT stars 
(Fig.~\ref{skydist}), 
preselected as nearby candidates on the
basis of their optical-to-infrared colours.
Some 30 of the southern nearby star candidates were detected in
a new high proper motion survey based on APM measurements 
(Scholz et al.\ \cite{scholz00}) and on the systematic search 
(described by Lodieu et al.\ \cite{lodieu05}) in
the SuperCOSMOS Sky Surveys
(SSS; Hambly et al.\ \cite{hambly01}) and 
in the 2MASS
but later identified with known NLTT stars.
Spectroscopic distances of 
non-NLTT stars, including high proper motion stars from other catalogues
and from 
the new
high proper motion survey in the southern sky, 
will be subject of the forthcoming paper in the series 
(paper IV, Scholz, Meusinger \& Jahrei{\ss} in preparation).

This paper is organised as follows: \S~\ref{hpm} describes the
selection of nearby NLTT candidates, their cross-identification
with 2MASS and preliminary distance estimates based on their 
optical-to-infrared colours.  \S~\ref{clspec} presents optical 
classification spectroscopy mainly obtained with the 2.2m telescope
at Calar Alto, including a comparison with previously available
spectroscopy of objects in our sample.  \S~\ref{spdist} shows the
main results of the study, the spectroscopic distance estimates 
and their statistics. In \S~\ref{indobj} we consider interesting individual
objects, and in \S~\ref{concdisc} we draw some conclusions.

%

\section{Selection of nearby NLTT candidates}
\label{hpm}

\subsection{Cross-
identification
of NLTT and 2MASS}
\label{xid}

When we started to combine the NLTT catalogue with the 2MASS data base
in early 2001, the revised NLTT catalogue with improved coordinates 
(Gould \& Salim \cite{gould03}; Salim \& Gould \cite{salim03}) 
was not yet available. The 2MASS was also not yet finished at that time,
and it was the 
2MASS Second Incremental Data Release, 
which we
used for defining our sample. 
On the one hand, the NLTT catalogue was available with highly uncertain
positions of high proper motion stars, but with poor optical photometry.
On the other hand, 2MASS has produced a catalogue with both high-accuracy 
near-infrared (NIR) photometry and accurate positions at epoch around 
1997-2000, but lacking proper motions.

The 2MASS data base provides a cross-identification with
optical data, i.e. the closest match within 5~arcsec with an object
from the Tycho-2 catalogue (H{\o}g et al.\ \cite{hog00}) or from the 
United States Naval Observatory (USNO) A2.0 catalogue 
(Monet et al.\ \cite{monet98}). The large epoch difference between the
2MASS observations and the USNO A2.0 catalogue leads to missing optical
counterparts of practically all high proper motion stars in the 2MASS
data base.

The most important criteria for the selection of $\sim$90\% of
the stars in our sample are described in the following three paragraphs
(i) to (iii). The remaining $\sim$10\% are briefly described in
\S~\ref{addsel}. Many candidates were actually selected according to
several of the selection criteria independently. 
Therefore, the 322 objects of our sample  (\S~\ref{finsample}) correspond
to about 500 candidates from the various initial candidate lists.

(i)
Since M dwarfs
are expected to be much brighter in the NIR than in
the optical, we first cross-identified the faint part of the 
NLTT ($m_\mathrm{r}>14.5$) 
with a sub-sample of bright ($K_\mathrm{s}<11$) 2MASS sources 
without optical counterparts (opt\_flg=0), 
extracted from the 2MASS data base (see also paper~I). 
A search radius of 60~arcsec was used taking into account the
large positional errors in the NLTT. From about 500 northern stars, 
initially selected in this way, 
about 70 candidates remained in our final sample according
to their preliminary distance estimates ($d<30$~pc) based on their 
$R-K_\mathrm{s}$ colours (see \S~\ref{selnear}).
A similar number ($\sim60$) of southern candidates were selected 
according to the above criteria after a cross-identification of
the whole southern NLTT with the same sub-sample of bright
2MASS sources without optical counterparts (see also paper~II). 

(ii)
As an alternative way to identify NLTT stars in the 2MASS, we used the 
Carlsberg Meridian Catalogs (CMC \cite{cmc99}), including a 
subsample of 12405 NLTT stars (CDS I/256/nltt), 
with accurate positions at epochs between 1984 and 1998.
The positions at epoch 2000, computed by 
VizieR\footnote{http://vizier.u-strasbg.fr/} taking into account the proper
motions, were then used for a direct identification with the 2MASS data base
using a smaller search radius of 10~arcsec. 
The information on the optical counterparts of the 2MASS sources
was not used here. The majority of the candidates in our final
sample ($>180$) was selected from these stars on the basis of
their preliminary distances below 25~pc obtained from their $(V-K)$
colours (see \S~\ref{selnear}).

(iii)
In addition to the NLTT sample in 
the CMC we also included stars with proper motions exceeding 100~mas/yr from 
the complete CMC 
(CDS I/256/stars)
and from Tycho-2 (H{\o}g et al.\ \cite{hog00}). 
The stars which turned out to be NLTT stars are analysed in this 
study. 
These are 120 CMC and 10 Tycho-2 stars, which were preliminarily
estimated to be within 25~pc based on their $(V-K)$
colours.
Other high proper motion stars, which were not found in the NLTT, 
are subject of paper IV (in preparation).
A small number of Tycho-2 and CMC high proper
motions could not be confirmed during the cross-identification process
(TYC~1597-481-1, TYC~3635-2220-1, TYC~2137-1529-1, CMC~1101791).

Using the CMC and Tycho-2 for the cross-identification with 2MASS did not
only give us the advantage of more accurate positions of the NLTT stars but
also provided better photometry (see section~\ref{optphot}). Since the bright
stars were expected to be measured by Hipparcos (ESA \cite{esa97}), we
identified only NLTT stars with $V>9$ (CMC) or $V_\mathrm{T}>9$ 
(Tycho-2) with 2MASS.

\begin{figure}[htb]
\begin{center}
\includegraphics[width=6.3cm,angle=0]{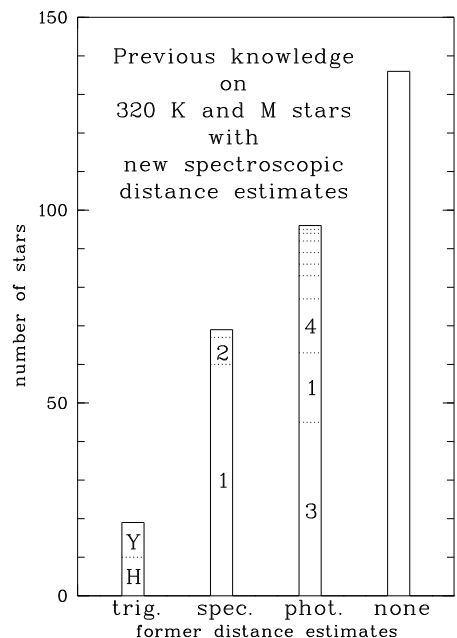}
\caption[numbers with d]{Numbers of objects in our sample with previous
distance estimates. Trigonometric distances are from the Hipparcos (H) and
Yale (Y) catalogues, ESA (\cite{esa97}) and van Altena, Lee \& Hoffleit
(\cite{vanaltena95}),
respectively. Most of the former spectroscopic and photometric distances
were derived by Reid and coworkers (1 = Reid et al.\ \cite{reid03b}; 2 =
Cruz \& Reid \cite{cruz02}; 3 = Reid, Kilkenny \& Cruz \cite{reid02b};
4 = Reid \& Cruz \cite{reid02a}).
}
\label{numbd}
\end{center} 
\end{figure}

Objects with well established distance estimates from Hipparcos 
(ESA \cite{esa97}) or ground-based trigonometric parallax measurements
were excluded, except in some doubtful cases or if they were lacking 
spectroscopy.  Objects with known spectral types (at the time when
our sample was defined - in 2001) were also excluded from our
target list for follow-up spectroscopy. As seen below (Fig.~\ref{numbd}),
about 20 objects with trigonometric distances remained in our sample,
whereas about one quarter of our objects got independent spectroscopic 
distance estimates in the course of our study. 

%

\subsection{Improved optical photometry}
\label{optphot}

Although the crude $m_\mathrm{r}$ magnitudes from the NLTT were used for part of our
objects in the early sample definition phase, our preliminary photometric 
distance estimates of nearby candidates were based on improved photometry. 
First, we used $R$ magnitudes from the USNO A2.0 catalogue (Monet 
et al.\ \cite{monet98}) in the northern sky, as described in paper~I. 
For southern sky objects (see also paper~II) we preferred to use 
SSS $R$ magnitudes (Hambly et al.\ \cite{hambly01}). The $R$ magnitudes
were assigned to the NLTT stars after a visual check of the NLTT/2MASS
identifications with the help of Digitized Sky Survey (DSS) images combined
with a catalogue query from the USNO A2.0 or with SSS images and catalogue
data. The SSS web site\footnote{http://www-wfau.roe.ac.uk/sss/}, providing 
both multi-epoch and multi-colour photographic
data in the form of images and object catalogues, and the DSS batch 
interface available at the National Astronomical Observatory of 
Japan\footnote{http://dss.mtk.nao.ac.jp/} were extremely helpful during
this process. 

More accurate $V$ magnitudes were obtained by the help of the already
mentioned VizierR interface from the CMC (\cite{cmc99}), the 
TASS Mark III photometric survey of the celestial equator 
(Richmond et al.\ \cite{richmond00}), and from
observations in the Geneva photometric system (Rufener~\cite{rufener88}),
if available. For the bright Tycho-2 stars in our sample we also
extracted the Tycho $V_\mathrm{T}$ (ESA \cite{esa97}) magnitudes.
There are some uncertain $V$ magnitudes in the CMC, as one can
find out from the CMC notes: for some objects without magnitude determination,
the magnitude was either taken (to one decimal place) from an input catalogue,
or it was assigned a nominal magnitude of 12. 

We did not apply the colour-dependent corrections to the CMC $V$ and Tycho-2
$V_\mathrm{T}$ magnitudes proposed in the original catalogues descriptions,
but
just used them for a rough photometric distance estimate, in order to select
the most promising candidates for spectroscopy. When using the photographic
$R$ magnitudes for our preliminary photometric distance estimates, we 
neglected possible systematic differences between the USNO A2.0 and SSS 
photometric systems.  Generally, we expected the
photographic $R$ magnitudes from the USNO A2.0 and from the SSS to have an 
accuracy of a few tenths magnitudes, whereas the $V$ magnitudes from various 
sources were probably a bit more accurate (typically $<\pm0.1$~mag).

%
\begin{figure}[htb]
\begin{center}
\includegraphics[width=6.4cm, angle=270]{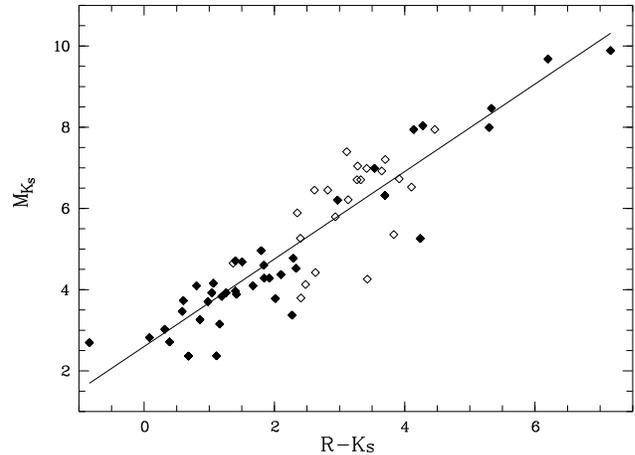}
\caption[Abs mag vs colour]{Absolute magnitude $M_{K_\mathrm{s}}$ 
as a function of $R-K_\mathrm{s}$ colour ($R$ magnitudes from 
SSS, $K_\mathrm{s}$ magnitudes from 2MASS).
Objects with trigonometric parallax errors $<$4~mas are shown as filled
symbols, those with larger errors - as open symbols.
The linear fit (solid line), based on the higher accurate sub-sample,
was used for estimating preliminary photometric
distances in order to select candidates within 30~pc for follow-up
spectroscopy.}
\label{MkR-K}
\end{center}
\end{figure}

\subsection{Selection based on optical-to-NIR colours}
\label{selnear}
 
In order to get first estimates on the distances of the NLTT stars
re-identified in archival optical data and cross-identified with the
infrared data from the second incremental release of the 2MASS,
we selected stars with known trigonometric parallaxes and computed
their absolute NIR magnitudes as well as their optical-to-infrared
colours. For all our candidates photographic $R$ magnitudes
were available either from the USNO A2.0 or from the SSS. A long colour
base line was expected to provide a strong correlation with absolute
magnitudes. Therefore, the $R-K_\mathrm{s}$ colours 
were chosen to be implemented in our preliminary distance estimates. 

Figure~\ref{MkR-K} shows the linear relationship
\begin{equation}
\label{preld}
M_{K_\mathrm{s}} = 2.60 + 1.077(R-K_\mathrm{s}),
\end{equation}
which we derived from using known nearby stars among the candidates found
in our SSS/2MASS cross-identification (see also paper~II). Only stars with 
trigonometric parallax errors below 4~mas (filled symbols) were used for the
fit. Other stars with larger errors are also plotted 
(open symbols). 
The spectral types, available for most of the stars shown 
in the figure, fall in the range between F7 and M7. 
The large scatter is mainly due to the
uncertainties of the photographic $R$ magnitudes. We found this calibration
satisfactory for our purpose. For simplicity, we applied the same
relationship for objects with $R$ magnitudes from the USNO A2.0 catalogue.

For objects with available $V$ magnitudes from the CMC (\cite{cmc99}), 
the TASS Mark III photometric survey (Richmond et al.\ \cite{richmond00}), or
from Rufener (\cite{rufener88}) we used the 
sixth-order polynomial relation between the absolute magnitude $M_\mathrm{V}$
and the $(V-K)$ colour derived by Reid \& Cruz \cite{reid02a} (see
their Figure~9 and the corresponding formula),
where the difference between $K$ and $K_\mathrm{s}$ magnitudes was neglected.
If only Tycho $V_\mathrm{T}$ magnitudes were available, 
we adopted  $V=V_\mathrm{T}$.

Objects were selected for follow-up spectroscopy if their rough photometric
distance estimates were below 25~pc or 30~pc based on the $V$ or $R$ 
magnitudes, respectively. A larger photometric distance limit was chosen when
the estimate was based on $R$ magnitudes, in order to take into account their 
larger uncertainties. Many objects were included in the target list according
to both selection criteria, and many objects appeared several times in the
final list as coming from different initial candidate lists (NLTT/2MASS plus
USNO A2.0 or SSS, CMC, Tycho-2, own high proper motion search using SSS).

\subsection{Additional NLTT stars in our sample}
\label{addsel}

About 30 stars were detected during various stages of a new 
high proper motion survey using first APM (Scholz et al.\ \cite{scholz00})
and later SSS measurements (see Lodieu et al.\ \cite{lodieu05}).
These stars were then identified as already known NLTT stars.
Most of them were also found as candidates in our main selection 
procedures described in the previous subsections. 

Another ten
known proper motion stars, mainly from the LHS catalogue,
were selected on the basis of their optical photometry
(from APM measurements)
and/or due to their uncertain spectral types 
(Luyten's rough ''m'' or ''k'' classification)
during the early stages of our survey, when 2MASS 
data were not yet available 
Further ten 
poorly investigated LHS stars, without
references in the Simbad data base 
bibliography\footnote{http://simbad.u-strasbg.fr/Simbad} were also
included. 
For the majority of these preselected $\sim$20 LHS stars
no good preliminary distances were available.
Many of these stars turned out to be rather distant,
according to our later spectral classification.

Most of the red dwarf stars in the CNS3 have spectral types from
Reid, Hawley \& Gizis (\cite{reid95}) and Hawley, Gizis \& Reid 
(\cite{hawley96}). 
A handful of additional
nearby candidates among the high proper 
motion stars omitted in 1991 from CNS3 due to poor or
missing optical photometry, and 
three more recently announced new 
nearby NLTT stars from  Fleming (\cite{fleming98})
were put on our observing program.  
Fleming's paper is a first 
attempt to fill the low velocity tail of the nearby star sample in
selecting stars on the basis of their X-ray flux and providing distances 
on subsequently obtained $VI$-photometry. Nevertheless, a large portion 
of Fleming's stars turned out to show significant proper motions. 
The low proper motion stars from Fleming's paper will be presented 
in paper IV. Last, but not least, also a 
some
of Hipparcos stars were included  
lacking
spectral types.
All ten Hipparcos stars contained in our sample were also selected  
from the CMC/2MASS cross-identification. Six of them had no previously  
known spectral types.

\subsection{The final sample}
\label{finsample}

Table~\ref{20obj}, available in its full length only via the 
Centre de Donn\'{e}es astronomiques de Strasbourg (CDS), lists
 all relevant information on the 322 spectroscopically classified
objects of our sample. Column 1 gives the object names, columns 2-7
list the J2000 coordinates taken from 
2MASS, 
column 8 gives the epoch of
the 2MASS observations, columns 9-11 list the 2MASS $JHK_\mathrm{s}$ 
magnitudes,
column 12 lists previously known spectral types, corresponding references
are in column 13, column 14 lists previously determined distances with their
references in column 15, column 16 lists the spectral types determined in
the present paper, with the corresponding spectroscopic distances in 
column 16, and the resulting tangential velocities
(using the original NLTT proper motions)
in column 17.

Designations are from the NLTT
(usually LP numbers, but also some Durchmusterung numbers), 
except for those starting with capital G, 
which were originally published in the
Lowell proper motion survey (Giclas, Burnham \& Thomas~\cite{giclas71}).
Proper motions are not listed, but are generally between 0.18 and 0.5
arcsec/yr. For 43 stars with proper motions larger than about
0.5 arcsec/yr we give the LHS number instead of the NLTT name.
For LHS~2288 and LHS~1200, which were below the
detection limit of 2MASS, the coordinates are taken from the latest
available SSS epoch measurements.
All references for previously known spectroscopy and/or distance 
determinations are given in the notes to Table~\ref{20obj}.

\begin{table*}
 \scriptsize
 \caption[]{Data on the first 20 stars out of the sample of 322. The
full table is available via the CDS. 
Distances are accurate to about 20\%.
}
\label{20obj}
 \begin{tabular}{lccrrrrrrrcrr}
 \hline 
 \hline
Name & $\alpha,\delta$ (J2000) & epoch & $J$ & $H$ & $K_\mathrm{s}$ & prev. & Ref. & prev. & Ref. & SpT & $d_{\rm spec}$ & $v_\mathrm{t}$ \\
(NLTT, LHS &  \multispan{5}{\hfil ~~------------------------------------(2MASS)--------------------------------------- \hfil} & SpT &  & dist. &  & \multispan{3}{\hfil ~~---------(this study)------------ \hfil}  \\
or Giclas)&                         &       &     &     &       &       &      & [pc]  &      & & [pc]       & [km/s] \\
 \hline
-17:6862  & 00 01 25.81 $-$16 56 54.2 & 1998.901 &  8.017 &  7.408 &  7.217 & M0.0 & 11 & 31.9 & 18 &  K7.0 &  30.2 &  50 \\
191- 43   & 00 08 55.13 $+$49 18 56.1 & 1998.849 & 10.864 & 10.320 &  9.980 & M5.5 & 26 & 15.8 & 26 &  M5.5 &  16.8 &  32 \\
644- 94   & 00 09 13.55 $-$04 08 02.0 & 1998.712 &  8.584 &  7.977 &  7.725 &      &    &      &    &  M2.5 &  20.3 &  20 \\
LHS~105    & 00 09 17.22 $-$19 42 32.4 & 1998.619 & 10.883 & 10.327 & 10.074 &      &    &      &    &  M4.0 &  31.8 & 174 \\
584- 94   & 00 17 40.65 $-$01 22 40.6 & 1998.707 &  9.237 &  8.567 &  8.356 &      &    & 28.6 & 26 &  M3.0 &  23.6 &  39 \\
705- 15   & 00 21 39.43 $-$09 00 24.2 & 1998.792 &  9.714 &  9.149 &  8.867 & M5.5 & 12 & 22.5 & 26 &  M3.5 &  24.0 &  34 \\
G158-073$^*$  & 00 24 25.20 $-$12 17 24.5 & 1998.605 &  8.825 &  8.213 &  7.979 &      &    & 26.3 & 26 &  M2.5 &  22.6 &  30 \\
G158-074$^*$  & 00 24 26.20 $-$12 17 19.8 & 1998.605 & 10.748 & 10.167 &  9.906 &      &    &      &    &  M4.0 &  29.9 &  39 \\
705- 28   & 00 25 50.97 $-$09 57 39.8 & 1998.819 &  9.883 &  9.250 &  8.952 &      &    &      &    &  M3.0 &  31.8 &  30 \\
193-488   & 00 26 02.59 $+$39 47 23.6 & 2000.441 & 10.990 & 10.398 & 10.102 & M4.5 & 26 & 30.8 & 26 &  M4.5 &  26.8 &  30 \\
G132-004  & 00 33 20.48 $+$36 50 26.2 & 1998.926 &  8.768 &  8.085 &  7.849 &      &    &      &    &  M3.0 &  19.0 &  43 \\
645- 48   & 00 34 38.96 $-$02 25 00.4 & 1998.710 &  8.574 &  7.951 &  7.692 &      &    & 24.1 & 26 &  M2.0 &  21.7 &  22 \\
705- 65   & 00 35 38.09 $-$10 04 18.6 & 1998.825 &  8.327 &  7.736 &  7.483 &      &    & 20.3 & 26 &  M2.5 &  18.0 &  17 \\
645- 53   & 00 35 44.13 $-$05 41 10.2 & 1998.710 & 10.667 & 10.074 &  9.711 & M5.0 & 13 & 18.8 & 13 &  M5.0 &  18.8 &  21 \\
585- 73   & 00 40 26.43 $-$00 08 40.9 & 1998.710 &  8.926 &  8.333 &  8.019 &      &    &      &    &  M2.5 &  23.7 &  25 \\
G270-100  & 00 56 30.21 $-$04 25 15.3 & 1998.704 & 10.444 &  9.832 &  9.563 &      &    & 30.0 & 15 &  M4.0 &  26.0 &  51 \\
LHS~1200   & 01 08 48.02 $+$15 15 22.7 & 1993.771 &        &        &        &      &    &      &    & WD    &  110.0 &  290 \\
G070-041  & 01 08 50.42 $-$03 40 33.7 & 1998.715 &  9.517 &  8.894 &  8.618 &      &    &      &    &  M3.0 &  26.9 &  37 \\
LHS~1204   & 01 09 31.25 $+$21 37 43.9 & 1997.778 & 10.087 &  9.534 &  9.272 & M1.0 & 16 & 27.8 & 42 &  M4.0 &  22.1 &  58 \\
G270-158  & 01 09 38.75 $-$07 10 49.7 & 2000.732 &  7.956 &  7.376 &  7.114 & M0.0 & 17 & 37.6 & 18 &  M2.0 &  16.3 &  32 \\ 
 \hline
 \end{tabular}

Notes:\\ 
Names flagged by $^*$ indicate that additional notes on these stars
are available in a separate notes file supplementing the electronic
table available at the CDS.
Some of the stars, for which their classification as normal K and M
dwarfs leads to extremely large tangential velocities, may in fact be
subdwarfs with roughly half as large distances and tangential 
velocities (see \S~\ref{sddist}).
References for previous spectral types and/or distance estimates are:
11 -- Hawley, Gizis \& Reid (\cite{hawley96}),
12 -- Gigoyan, Hambaryan \& Azzopardi (\cite{gigoyan98}),
13 -- Cruz \& Reid (\cite{cruz02}),
14 -- Reid \& Cruz (\cite{reid02a}),
15 -- van Altena, Lee \& Hoffleit (\cite{vanaltena95}),
16 -- Buscombe (\cite{buscombe98}),
17 -- Turon et al.\ (\cite{turon93}),
18 -- ESA (\cite{esa97}),
22 -- Reid, Kilkenny \& Cruz (\cite{reid02b}),
24 -- R\"oser \& Bastian (\cite{roeser88}),
26 -- Reid et al.\ (\cite{reid03b}),
27 -- Weis (\cite{weis86}),
28 -- Torres et al.\ (\cite{torres00}),
29 -- Kirkpatrick, Henry \& Simons (\cite{kirkpatrick95}),
30 -- Tokovinin (\cite{tokovinin94}),
32 -- Weis (\cite{weis88}),
33 -- Weis (\cite{weis87}),
36 -- Gliese \& Jahrei{\ss} (\cite{gliese91}),
37 -- Gizis (\cite{gizis00}),
39 -- Lee (\cite{lee84}),
42 -- Weis (\cite{weis84}),
43 -- Li \& Hu (\cite{li98}),
44 -- Weistrop (\cite{weistrop81}),
45 -- Fleming (\cite{fleming98}),
46 -- paper I.

\end{table*}
 
\normalsize

\section{Follow-up classification spectroscopy}
\label{clspec}

\subsection{Spectroscopic observations and data reduction}
\label{obsred}

Optical spectroscopic data were collected over a period of five years. With 
only a few exceptions, all objects from our sample were observed with the 
2.2\,m telescope of the German-Spanish Astronomical Centre on Calar Alto, 
Spain, i.e. using the same instrument setup as in paper~I. 
The focal reducer and faint object spectrograph CAFOS was used with 
the grism B\,400 giving a wide wavelength coverage from about 3\,600\,$\AA$\, 
to 8\,500\,$\AA$ and a dispersion of 10\,$\AA$\, per pixel on the SITe1d CCD.
With a width of the entrance slit
of typically between 1 and 2\,arcsec (dependent on the seeing conditions),
the effective resolution is about 20 to 30\,$\AA$\, FWHM.
The vast majority of our objects were observed during two dedicated
observing runs with {\it (a)} 10 nights in March 2002 (visitor mode) and
{\it (b)} 6 nights in August/September 2002 (service mode). For the other
targets spectra were taken in the frame of backup programs for 10 different
observing runs with CAFOS during the period from 1999 to 2004.
 
One star with missing CAFOS spectrum (G008-017) as well as LHS~1200, for
which we needed a higher-resolution spectrum for a radial velocity
measurement (\S~\ref{wddist1}), were observed with the
Nasmyth Focal Reducer Spectrograph at the 2\,m telescope of the Th\"uringer
Landessternwarte (TLS) Tautenburg in February 2004 and January 2005, 
respectively. The V\,200 grism was used which yields a dispersion of
3.4\,$\AA$\, per pixel, or 12\,$\AA$\, FWHM for a slit width of 1\,arcsec.
Finally, for the star LP~869-26, which had a corrupted CAFOS spectrum, an
additional spectrum was taken with the low-resolution spectrograph DOLORES
at the 3.6\,m Telescopio Nazionale Galileo (TNG) on the island of La Palma,
Spain in August 2004. The grism LR-R (2.9\,$\AA$\, per pixel) was used;
the resolution of the spectrum is  20\,$\AA$\, FWHM
(see Fig.~\ref{11later}).

A large number of comparison stars of known spectral types, mainly taken
from the NStars data base\footnote{http://nstars.arc.nasa.gov/} and
from the ARICNS data base for nearby 
stars\footnote{http://www.ari.uni-heidelberg.de/aricns/} were observed
during the nights of our target observations. In the range of spectral 
types between M0.0 and M6.5 such comparison stars were selected for nearly
all spectral sub-types, whereas the coverage for K and late-M dwarfs was
less complete. Observations of comparison stars were repeated regularly,
with some spectral types being represented by several template stars.
During the observing nights at the TLS 2\,m telescope and at the TNG, 
only one or two comparison stars, with spectral types close to those of 
the targets, were included.
 
All spectra were reduced with standard routines from the ESO MIDAS data
reduction package. Wavelength calibration was done with spectra of
calibration lamps, except for the TLS spectra for which
the night-sky lines were used following Osterbrock \& Martel (\cite{Ost92}).
Flux-calibration was done by means of standard
stars
from Oke (\cite{oke90}).  
The one-dimensional spectra were extracted from the two-dimensional
spectra using both the optimal extraction algorithm by Horne (\cite{horne86})
and a simple extraction method. 
In most cases, the results from both methods agree very well.
For the faintest stars, the Horne algorithm provides a better
signal-to-noise ratio. On the other hand, this method is more 
sensitive to variations of the background flux (e.g., due to 
stray light) and breaks down in the presence of strong background 
gradients. In these few cases, the results from the simple 
extraction were used.

\subsection{Classification of M- and K-type spectra}
\label{clsMK}

\begin{figure}[htb]
\begin{center}
\includegraphics[width=6.4cm, angle=270]{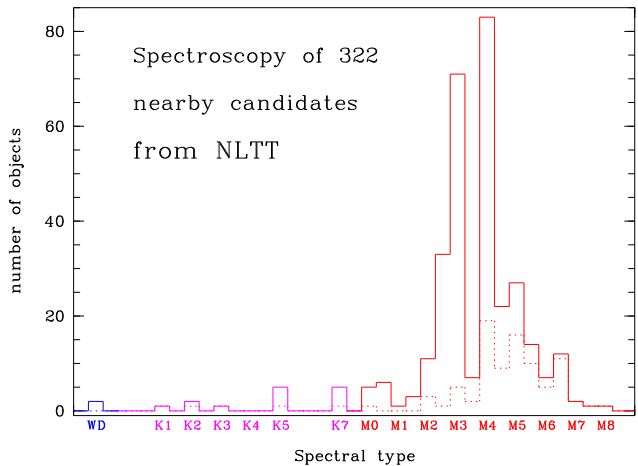}
\caption[SpT histogram]{Distribution of 322 objects with their spectral types
determined in this study (solid line). The dotted line shows the 
subsample of 90 objects, for which we
found previously published spectroscopy (see also Fig.\ \ref{newSpTref}).
}
\label{SpThis}
\end{center}
\end{figure}

The large number of comparison stars with known spectral types, 
spectroscopically observed together with our target stars, served as the
basis for our spectral classification.
These template spectra, shown in Fig.~\ref{bluestd} and Fig.~\ref{redstd},
represent in many cases the mean of several standard spectra of a given
spectral type. As one can see, the template spectra form a sequence, which
is as expected, sorted by spectral type 
(Figs.~\ref{bluestd} and~\ref{redstd}). The only exception is the star
Gl~273, which was planned to be used as the M3.5 standard. 
Since this star turned out to have quite a peculiar
 spectrum, resembling an M2.5 star in the blue part
(Fig.~\ref{bluestd}) but exhibiting a much too red continuum in the red
(Fig.~\ref{redstd}), there were essentially no target stars with comparable
spectra. A few target spectra, which were just in between M3 and M4, were 
however classified as M3.5 type. Therefore, the histogram of the spectral 
types of our targets (Fig.~\ref{SpThis}) shows an apparent deep minimum for 
M3.5 stars. We 
assume
that $\sim$30\% of our M3 and M4 dwarfs 
could be in fact M3.5 dwarfs.

The classification based on the comparison with template spectra worked as
follows: first, both the template and the target spectra were normalised
at 7500\AA{}. Then we computed the absolute 
values of the differences between the flux densities of the target and the 
template, 
$\delta(\lambda) = | f_{\rm targ}(\lambda)-f_{\rm temp}(\lambda)|$,
For every comparison with a template $i$, the mean value 
$\bar{\delta_{\rm i}}$ and the median $\tilde{\delta_{\rm i}}$ were
computed. The spectral type of the target was subsequently identified 
with the spectral type of those templates where 
$\bar{\delta_{\rm i}}$ and $\tilde{\delta_{\rm i}}$
reach their minima. Usually, 
$\bar{\delta_{\rm i}}$ and $\tilde{\delta_{\rm i}}$
were at minimum with the same template. In cases, where 
$\bar{\delta_{\rm i}}$ and $\tilde{\delta_{\rm i}}$ led to different 
classifications, we simply adopted the spectral type corresponding 
to the smaller value of the two minima. 
In 
about 10
cases,
where the minimum mean value was similar to the minimum median value, but 
their corresponding template spectral types differed by more than half a
spectral sub-type, we took the average spectral type of these templates.

The classification of target spectra later than M6.5 or earlier than M0
required some interpolation and additional visual comparison of the spectra,
since fewer template spectral sub-types were available for K-type and late-M
spectra.
For all stars classified by the comparison with template spectra as
K-type, in particular for early-K stars, we used the atlas of
spectra by  Torres-Dodgen \& Weaver (\cite{torres93}) for an additional
check. Stars with H${\alpha}$ absorption lines stronger than their
NaI lines (see Figure~6 in Torres-Dodgen \& Weaver~\cite{torres93})
were classified as G-type and excluded from our sample of nearby candidates.
This concerned HD~206058 and LHS~3856. However, in case of LHS~3856 we
are not sure that the correct target was observed spectroscopically.

The few spectra not observed at Calar Alto but with the TLS 2\,m telescope 
and with the TNG, were classified by visual comparison with the few templates
observed for comparison at the same telescope. Fortunately, these comparison
stars were selected to have spectral types similar to our targets. However,
some extrapolation was needed in these few cases. Of course, the Calar Alto
templates could be used for an additional check of the results, which were
consistent.

We did also try to use spectral indices as described in the classification
scheme for M dwarfs developed by Reid, Hawley \& Gizis~(\cite{reid95}), in
particular the bandstrengths of TiO and CaH. However, the resolution of our
spectra was much too low, in order to achieve reliable results. Therefore,
we preferred our direct comparison with template spectra taken at the same
spectral resolution and with the same instrument.

The distribution of spectral types derived for the 322 NLTT stars of the 
present study shows a pronounced maximum between M2 and M4.5;  
the majority of the stars have spectral types earlier than M5
(Fig.\ \ref{SpThis}).  There is an apparent minimum at spectral 
type M3.5 which is  
an artifact, 
caused by the fact that
our M3.5 reference star, Gl~273, turned out to be not useful as a 
template (see above).

\subsection{Comparison with other spectroscopy}
\label{prevspec}

Meanwhile, published spectral types are available for about
one quarter of the objects in our sample, mainly from the recently 
completed survey by Reid et al.\ (\cite{reid03b}) and Cruz \& Reid 
(\cite{cruz02}). For this subsample, the fraction 
of late-type ($>$M5) M dwarfs is higher than for our whole NLTT sample.
Consequently, the spectral types newly added by the present study 
are strongly concentrated at early-M dwarfs.

%
 
\begin{figure}[htb]
\begin{center}
\includegraphics[width=6.3cm,angle=270]{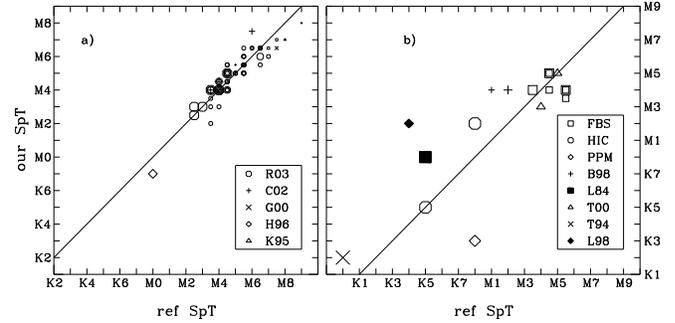}
\caption[New and Ref SpT]{The new spectral types derived in this study
compared to previous spectroscopy - a) for 72 objects from
R03: Reid et al.\ (\cite{reid03b}),
C02: Cruz \& Reid (\cite{cruz02}),
G00: Gizis (\cite{gizis00}),
H96: Hawley, Gizis \& Reid (\cite{hawley96}),
K95: Kirkpatrick, Henry \& Simons (\cite{kirkpatrick95}),
and b) for 18 objects from
FBS: Gigoyan, Hambaryan \& Azzopardi (\cite{gigoyan98}),
HIC: Hipparcos Input Catalogue (Turon et al.\ \cite{turon93}),
PPM: Positions and Proper Motions - North catalogue
     (R\"oser \& Bastian \cite{roeser88}),
B98: Buscombe (\cite{buscombe98}),
L84: Lee (\cite{lee84}),
T00: Torres et al.\ (\cite{torres00}),
T94: Tokovinin (\cite{tokovinin94}),
L98: Li \& Hu (\cite{li98}). One object with G5 spectral type from HIC,
classified by us as a K1 star, lies outside of the plotted range.
The symbol sizes correspond to the magnitudes of the objects,
with larger symbols used for brighter objects.
Solid lines represent the coincidence of new and previous
spectral types.
}
\label{newSpTref}
\end{center}
\end{figure}

The majority (60) of the objects with previous spectroscopy come from the
work by Reid et al.\ (\cite{reid03b}). The mean difference between our and
their spectral types is exactly zero with a dispersion of 0.50 subtypes. 
If we add 12 objects from other spectroscopic studies, which we assume to
be in the same system (Fig.\ \ref{newSpTref}a), we get a mean difference
of $-$0.01 with a dispersion of 0.56 subtypes. Other previously published
spectral types for objects in our sample are shown in Fig.\ \ref{newSpTref}b.
Among the latter we see larger discrepancies, in particular for stars
with earlier spectral types. The extreme outlier seen in Fig.\ \ref{newSpTref}b
is LP~247-13, for which we assigned a spectral type of M2.0, whereas 
Li \& Hu (\cite{li98}) classified it as K4. However, we note for this 
particular object a very blue continuum compared to the typical 
early M dwarf spectrum. A similar spectrum was observed by us for LHS~6388. 
These two objects may have blue companions, unresolved in our observations.

%

The relatively low resolution of our spectroscopy combined with
the direct comparison of the full spectra with templates provides
a classification nearly as accurate as achieved
by others using slightly higher resolution classification
spectroscopy in the same
system (see Fig.~\ref{newSpTref}a). Compared to other previous
classifications (Fig.~\ref{newSpTref}b), often not corresponding 
to the presently adopted system, our spectral types should be prefered.

\subsection{
Two
White dwarf spectra}
\label{wdspec}

Two objects in our sample turned out to be white dwarfs. These are LHS~1200,
which was originally selected due to its uncertain spectral type, and 
LHS~2288, which we selected as a faint red object ($E=18.7, O-E\sim+3.2$)
according to the APM catalogues (McMahon, Irwin \& Maddox \cite{mcmahon00}).
The SSS magnitudes of LHS~1200 and LHS~2288 
are $B_\mathrm{J}=16.92, R=16.79$ and $B_\mathrm{J}=20.21, R=18.50$, 
respectively. Both stars have almost completely been 
neglected in the literature, although LHS~1200 was included in the study
of Eggen (\cite{eggen68}) and is part of Luyten's white dwarf catalogue
(Luyten \cite{luyten70}). The two objects are below the detection limit
in the 2MASS. This is, with respect to their optical magnitudes and colours,
already indicative of their possible white dwarf nature. 

\begin{figure}[htb]
\begin{center}
\includegraphics[width=6.3cm,angle=270]{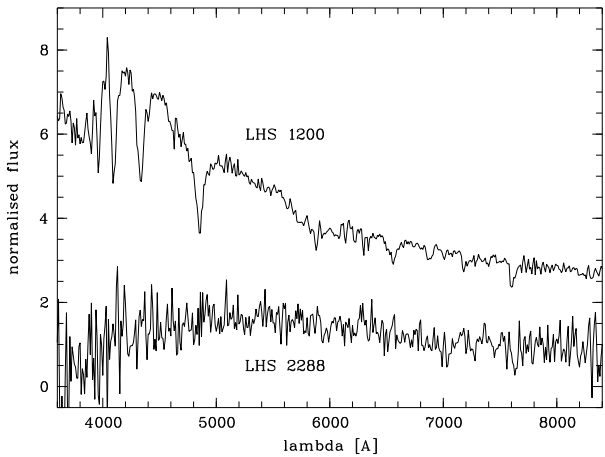}
\caption[two WD spectra]{CAFOS low-resolution spectra of two white dwarfs.
The spectra are normalised at 7500\AA{}, that of LHS~1200 is 
offset
by two flux units.
}
\label{2wd}
\end{center}
\end{figure}

Our low-resolution spectra of LHS~1200 and LHS~2288 are shown in 
Fig.~\ref{2wd}. LHS~1200 can be classified as an early-type
DA white dwarf with broad Balmer lines, similar to the spectrophotometric 
standards Wolf~1346 and GD~140 (Massey et al.\ \cite{massey88}). 
A more accurate classification of LHS~1200 will be given in \S~\ref{wddist1}
based on photometry. LHS~2288 appears with the given 
signal-to-noise to be a featureless cool DC white dwarf, similar to the
spectra shown in Fig.~2a of Oppenheimer et al.\ (\cite{oppenheimer01}).
Again, accurate photometry allows an estimate of the effective temperature
of this object (see \S~\ref{wddist2}).

%

\section{Spectroscopic distance estimates}
\label{spdist}

\subsection{K and M dwarfs}
\label{KMdist} 

Figure~\ref{numbd} shows that about 20 objects in our sample 
of 322 objects have
measured trigonometric parallaxes. All but 4 of these objects were previously
lacking spectral types. About 70 objects have recently determined
spectroscopic distances. Note that the total number of objects with previous
spectral types is 90 (cf. Fig.~\ref{SpThis} and Fig.~\ref{newSpTref}).

We wanted to be comparable with the spectroscopic distance estimates of
Cruz et al.\ (\cite{cruz03}) using their $M_\mathrm{J}$-spectral type 
relationship. However, this calibration is only valid for spectral types
between M6 and L8, i.e. with only little overlap to the range of spectral 
types in our sample. In order to expand this calibration to earlier spectral 
types, we carried
out a new calibration on the basis of all 
K, M and L dwarfs 
in the forthcoming CNS4 (Jahrei{\ss} \cite{jahreiss05})
with trigonometric parallaxes better than 20 per cent. The calibration was
performed in applying a robust locally weighted regression LOWESS as
described in Cleveland (\cite{cleveland79}), 
with a smoothing factor of $f = 0.3$.

\begin{figure}[htb]
\begin{center}
\includegraphics[width=6.3cm,angle=270]{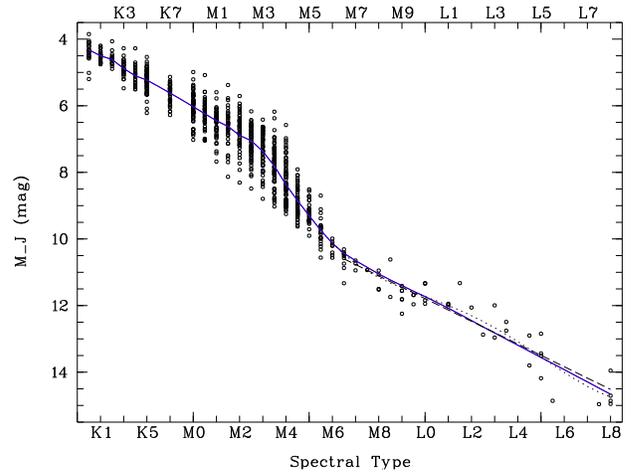}
\caption[abs J mag vs SpT]{Absolute magnitudes/spectral types calibration.
The solid line represents our new calibration. For comparison we show the
linear fit of Dahn et al.\ (\cite{dahn02}; dashed line) and the fourth-order
polynomial fit by Cruz et al.\ (\cite{cruz03}; dotted line), which are both
valid for later spectral types only ($>$M6).
Adopted mean $M_\mathrm{J}$ magnitudes are listed in Table~\ref{MJSpT}.
}
\label{absmagspt}
\end{center}
\end{figure}

Figure~\ref{absmagspt} shows our new calibration, which 
at later spectral types ($>$M8 - not relevant for the objects in our
sample) is very similar to a linear fit, and which over the whole overlapping
spectral range (M6-L8) lies in between the linear fit by  
Dahn et al.\ (\cite{dahn02}) and the higher-order fit by 
Cruz et al.\ (\cite{cruz03}). The overall shape of the fit could also roughly
be described by three linear relations in the intervals of K0-M3, M3-M6.5
and M6.5-L8, respectively. However, we prefer not to approximate the 
calibration by formulae, and list the relationship in Table~\ref{MJSpT}.
The dispersion is generally of the order of 0.4-0.5~mag.
This transforms to a distance error of $\sim$20\%.

%

\begin{table}
 \footnotesize
 \caption[]{Absolute magnitudes $M_\mathrm{J}$ 
as a function of spectral types between 
K0 and L8 adopted for the distance estimates of the objects in our sample.
}
\label{MJSpT}
 \begin{tabular}{rrrrrr}
 \hline 
 \hline
Sp.Type & $M_\mathrm{J}$ & Sp.Type & $M_\mathrm{J}$ & Sp.Type & $M_\mathrm{J}$ \\
 \hline
K0   &  4.32 & M0.0 &  6.03 & L0.0 & 11.73 \\
     &       & M0.5 &  6.24 & L0.5 & 11.90 \\
K1   &  4.50 & M1.0 &  6.45 & L1.0 & 12.08 \\
     &       & M1.5 &  6.62 & L1.5 & 12.26 \\
K2   &  4.61 & M2.0 &  6.89 & L2.0 & 12.45 \\
     &       & M2.5 &  7.05 & L2.5 & 12.64 \\
K3   &  4.88 & M3.0 &  7.37 & L3.0 & 12.82 \\
     &       & M3.5 &  7.81 & L3.5 & 13.01 \\
K4   &  5.10 & M4.0 &  8.37 & L4.0 & 13.19 \\
     &       & M4.5 &  8.85 & L4.5 & 13.37 \\
K5   &  5.23 & M5.0 &  9.30 & L5.0 & 13.56 \\
     &       & M5.5 &  9.74 & L5.5 & 13.74 \\
     &       & M6.0 & 10.13 & L6.0 & 13.92 \\
     &       & M6.5 & 10.43 & L6.5 & 14.11 \\
K7   &  5.62 & M7.0 & 10.65 & L7.0 & 14.29 \\
     &       & M7.5 & 10.85 & L7.5 & 14.48 \\
     &       & M8.0 & 11.05 & L8.0 & 14.66 \\
     &       & M8.5 & 11.23 \\
     &       & M9.0 & 11.40 \\
     &       & M9.5 & 11.57 \\
 \hline
 \end{tabular} 
\end{table} 

\begin{figure}[htb]
\begin{center}
\includegraphics[width=6.3cm,angle=0]{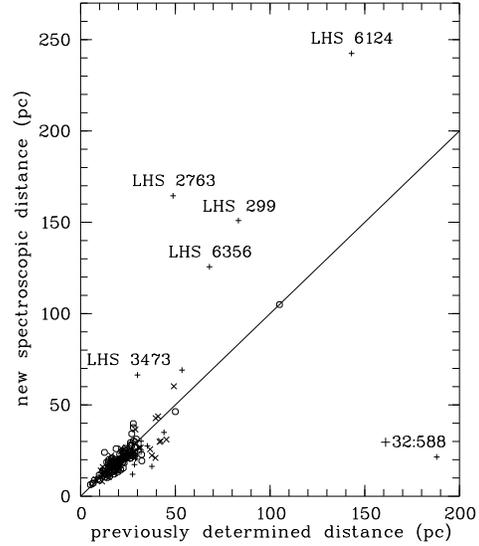}
\caption[distances comparison]{The newly determined spectroscopic distances 
of 166 objects in our sample compared to their
previous trigonometric ($+$), spectroscopic ($\times$) and photometric (open
circles) estimates. The line represents equal distances. The objects with
the largest discrepancies are labeled (see text).
}
\label{dvsd}
\end{center}
\end{figure}

Figure~\ref{dvsd} shows that our distance estimates generally agree well with 
former estimates if available from trigonometric or recent spectroscopic and
photometric measurements (see also Fig. \ref{numbd}). The object with the
largest previously determined photometric distance in our sample is  LHS~2048
($d_{\rm phot}=105$~pc; Weistrop \cite{weistrop81}), which we confirm by our
classification of this star as an M2.5 dwarf. However, there are some
objects with large discrepancies (labeled in Fig.~\ref{dvsd}), which need to
be explained. This seems even more important since in all the
problematic cases the stars have previous
trigonometric parallax measurements. Interestingly, all outliers above the
line have ground-based parallaxes, while the only one object far below the
line of equal distances has a Hipparcos parallax.

\begin{figure}[htb!]
\begin{center}
\includegraphics[width=9.0cm,angle=0]{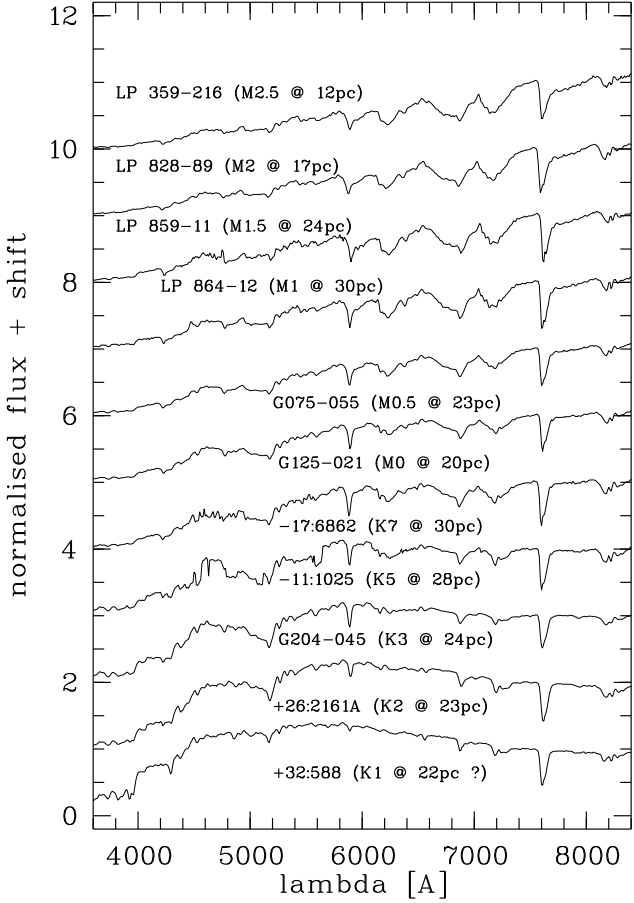}
\caption[nearby early-type obj]{CAFOS spectra of nearby early-type ($<$M3) 
objects.  
One of the nearest objects of each spectral type (according 
to our distance estimates) is shown.
The K1 star, BD$+$32$^{\circ}$588, has a Hipparcos parallax, 
which places this star nearly ten times farther away (see text).
}
\label{11early}
\end{center}
\end{figure}

The Hipparcos star BD$+$32$^{\circ}$588 (Hip 15102; ESA \cite{esa97}) is the only star in
our sample of K and M proper motion stars, which has a previous spectroscopic
classification as a G-type star (G5; Turon et al.\ \cite{turon93}). 
According to our spectrum taken at Calar Alto (spectrum at the bottom of 
Fig.~\ref{11early}), we classified it as a
K1 star, based on the comparison with template spectra and 
on the fact that the NaI absorption line is slightly stronger 
than the H$\alpha$ line 
(cf. Figure~6 in Torres-Dodgen \& Weaver~\cite{torres93}). Its $J-K=0.63$
is also more consistent with an early K-type rather than with G5. The star 
was suspected by Halliwell (\cite{halliwell79}) as a possible nearby star 
($d<25$~pc) but not further mentioned in the literature. The Hipparcos 
parallax of 5.32$\pm$1.19~mas is in strong contrast to our nearly ten times
smaller spectroscopic distance estimate of about 21~pc. 
The Hipparcos distance and 
the bright 2MASS magnitude $J=6.16$
are not consistent with a main sequence star. 
This confirms that this object is an early K giant with  
$M_V=1.6$ and $B-V=0.96$, clearly brighter than a
subgiant (with typical $M_V\sim3.2$).

There are five objects where available trigonometric parallaxes
show our spectroscopic distances to be too large:
LHS~6356 (K5), LHS~6124 (K7), LHS~2763 (M0), LHS~299 (M0),
and LHS~3473 (M3). 
These objects
labeled in Fig.~\ref{dvsd},
may in fact be subdwarfs (see \S\ref{sddist}). 
LHS~3473 and LHS~6356 have common proper motion companions
(see \S\ref{cpms}).

%

\begin{figure}[htb!]
\begin{center}
\includegraphics[width=9.0cm,angle=0]{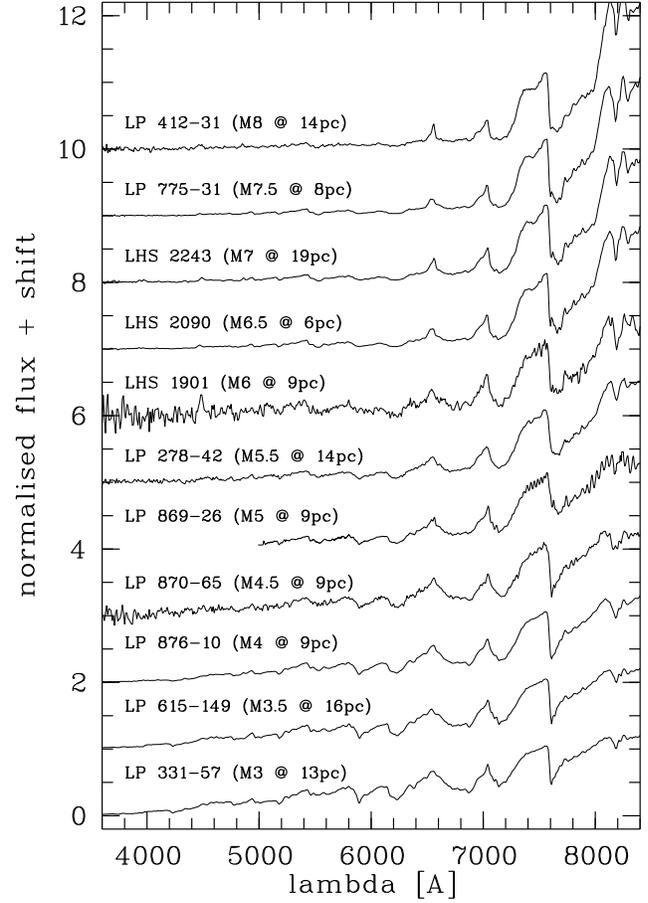}
\caption[nearby later-type obj]{Spectra of nearby mid- and late-type 
(M3--M8) objects.
One of the nearest objects of each spectral type (according
to our distance estimates) is shown.
All spectra were taken with
CAFOS, except for LP~869-26, which was obtained with the TNG.
}
\label{11later}
\end{center}
\end{figure}

Fig.~\ref{11early} and Fig.~\ref{11later} show representative samples
of our spectra of K and early M dwarfs and of mid- to late-type M dwarfs,
respectively.

%

%

\subsection{Subdwarf candidates}
\label{sddist} 

As mentioned already in Sect.\,\ref{clsMK},
our spectral resolution is not high enough for computing 
reliable TiO and CaH indices (Reid, Hawley \& Gizis \cite{reid95}) of our 
targets.  Therefore, we can not distinguish between normal K and M dwarfs 
and subdwarfs using these indices according to the classification scheme 
for subdwarfs developed by Gizis (\cite{gizis97}), 
which can in principle only be applied to M subdwarfs.
The spectral resolution is in fact about 7 times lower than in
Gizis (\cite{gizis97}). Therefore, the relatively sharp TiO features 
are considerably smoothed, leading to systematically too large TiO5
indices. The CaH2 and CaH3 indices are also affected by smoothing 
in the region of the maximum at 7042-7046\,$\AA$\, whereas the CaH1 
index does obviously make no sense with a resolution comparable to the
distances between the wavelength intervals used for its definition.

We do, however, expect some
subdwarfs in our proper motion selected sample. We have used the mean absolute
magnitudes of normal K and M dwarfs with trigonometric parallax measurements
for the spectroscopic distances of all objects in our sample. By doing so,
we overestimate the distances of those objects, which are in fact sub\-dwarfs.
The comparison of the spectroscopic distances of some of our objects with 
previous determinations from ground-based trigonometric parallax measurements 
has already revealed some discrepancies (Fig.\ \ref{dvsd}).

%
\begin{figure}[htb]
\begin{center}
\includegraphics[width=6.3cm,angle=270]{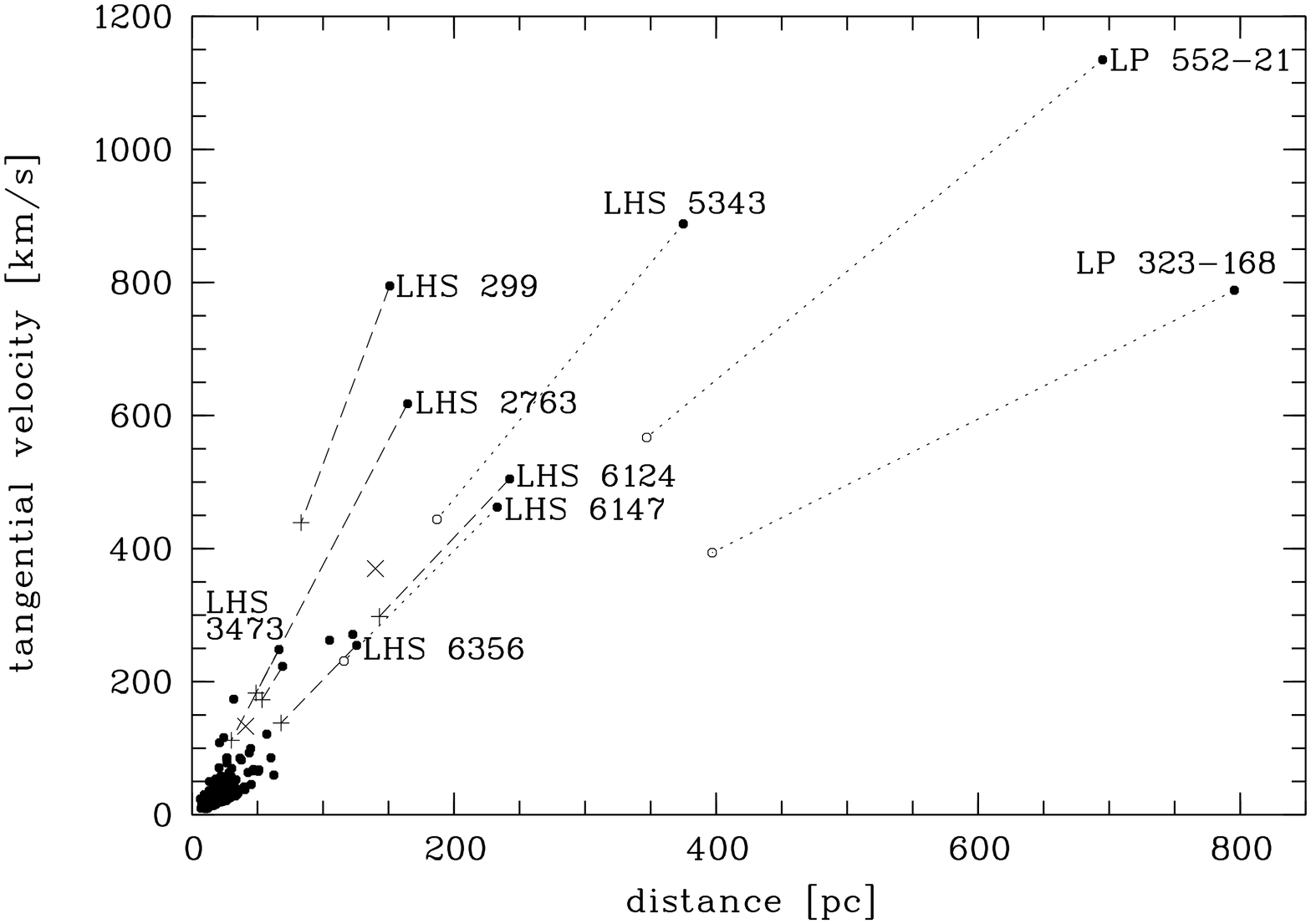}
\caption[vt over d]{Tangential velocities versus distances. The
spectroscopic distances and resulting tangential velocities of 320 K and M
dwarfs are shown by filled circles.
The two white dwarfs are marked as crosses.
Objects with trigonometric parallax data in addition to our spectroscopic
distances are shown by dashed lines and plus signs ($+$).
Candidate subdwarfs are illustrated by dotted lines and open circles.
If these are in fact subdwarfs, the
values derived from the spectroscopic distances are reduced by as
much as 50\%
}
\label{vtd}
\end{center}
\end{figure} 

Figure \ref{vtd} shows the spectroscopic distance estimates and the resulting
tangential velocities of all 322 objects in our sample. The four most extreme
objects as well as the five objects with already mentioned overestimated
distances (Fig. \ref{dvsd}) are labeled. Among the stars with trigonometric
parallaxes in our sample, LHS~299 has the largest tangential velocity of
about 440~km/s. Despite this previously known large velocity, there are no
hints in the literature on the subdwarf nature of this object. Our distance
estimate based on the assumed classification of LHS~299 as a normal M0 dwarf 
is nearly twice as large as its trigonometric distance. This discrepancy
can only be explained, if this object is in fact a subdwarf.

We expect that the distances to at least four other suspected subdwarfs 
may be only about half of our original estimates
(indicated by the dotted lines and open circles in Fig. \ref{vtd}).
Our original spectral classification of these four objects (LP~552-21, 
LP~323-168, LHS~5343, LHS~6147) ranges between K2 and M0.5, and for a first
guess we can assume similar subdwarf subtypes. There are four sdK7, three
sdM0.0 and three sdM0.5 stars with ground-based trigonometric parallaxes 
listed in Gizis (\cite{gizis97}). Using for some of these objects the
Hipparcos (ESA \cite{esa97}) parallaxes instead of the ground-based ones, 
we compute mean absolute magnitudes $M_\mathrm{J}$ of 6.9, 7.5 and 7.8 for 
sdK7, sdM0.0 and sdM0.5, respectively. Compared
to the absolute magnitudes of normal K7, M0.0 and M0.5 dwarfs, these values
are 1.3 to 1.6~mag fainter, corresponding to a factor of 0.55 to 0.48 in
the computed distances,
and thus reduce their tangential velocities by half.

\begin{figure}[htb]
\begin{center}
\includegraphics[width=9.0cm,angle=0]{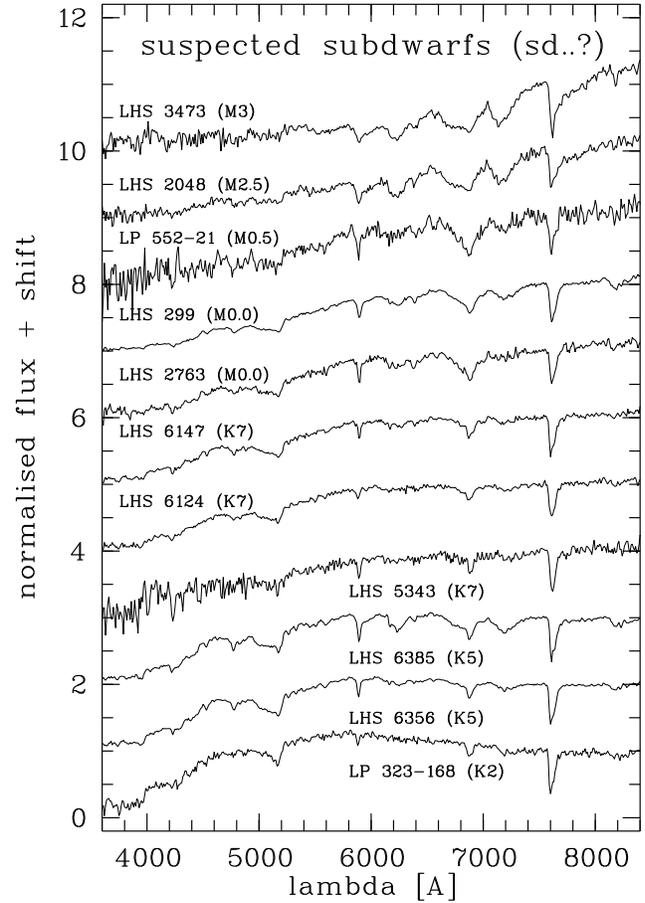}
\caption[high-velocity obj]{CAFOS spectra of all high-velocity K and M stars
(with approximately $v_\mathrm{t}>250$~km/s) 
according to our spectroscopic distance estimates,
assuming that these objects are normal dwarf stars.
All these objects
may in fact be subdwarfs. 
}
\label{11sd}
\end{center}
\end{figure}

\begin{figure*}[htb]
\begin{center}
\includegraphics[width=13.0cm, angle=270]{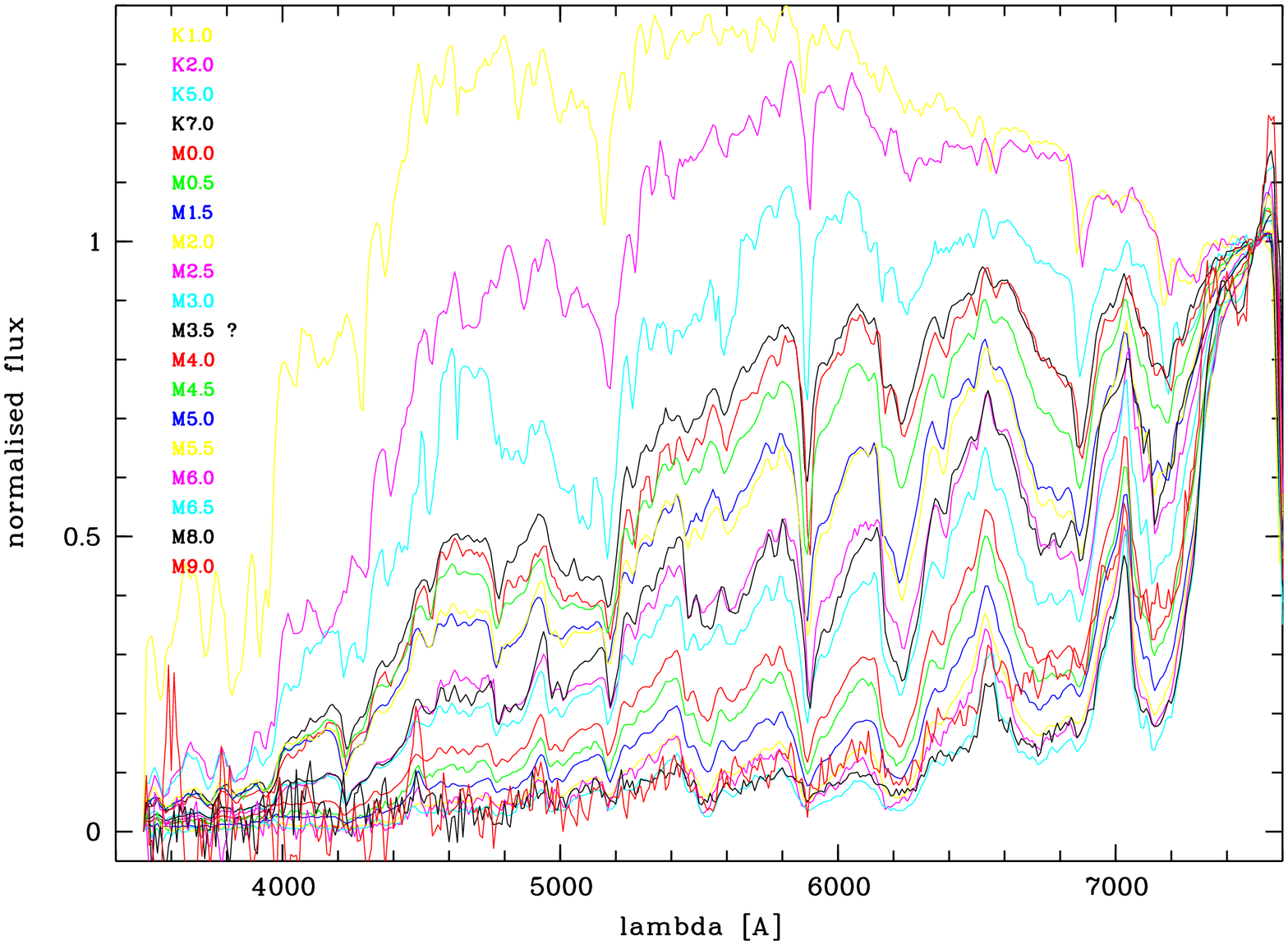}
\caption[Standards blue sp]{{\it Available only in online version.}
Blue part of the standard spectra, normalised at 7500\AA{},
used for the classification of the targets.
The spectra form a sequence with increasing spectral type. The only exception
is the M3.5 standard (Gl~273), which is more similar to M2.5 in the spectral
range shown.
}
\label{bluestd}
\end{center}
\end{figure*}

\begin{figure*}[htb]
\begin{center}
\includegraphics[width=13.0cm, angle=270]{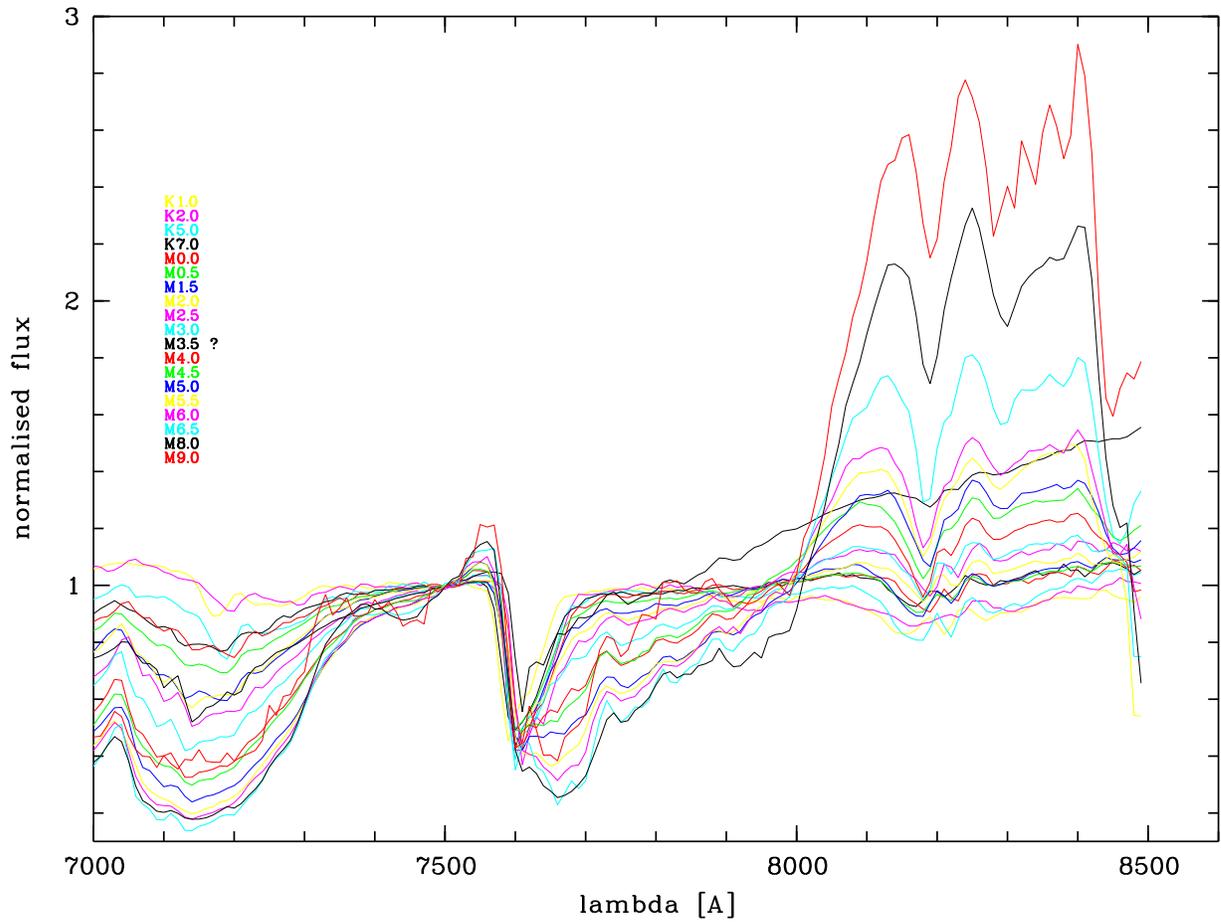}
\caption[Standards red sp]{{\it Available only in online version.}
Red part of the standard spectra, normalised at 7500\AA{},
used for the classification of the targets.
Again, the spectra form a sequence with increasing spectral type, except for
M3.5 (Gl~273), which appears much too red in this spectral range.
}
\label{redstd}
\end{center}
\end{figure*}

Figure~\ref{11sd} shows the spectra of all K and M stars in our sample which 
have the largest tangential velocities, from $\sim$250~km/s to
$\sim$1150~km/s (see also Fig.~\ref{vtd}), based on their classification 
as normal dwarfs and the resulting large distances. 
With the given low resolution, these spectra do not provide any evidence on
a subdwarf nature of the objects. The spectrum of the highest-velocity
object, LP~552-21, looks a bit peculiar but does not allow an alternative
spectral classification due to the low signal-to-noise. Comparison with
the spectra of local K and early-M dwarfs (cf. Fig.~\ref{11early}) does not
reveal any systematic differences.

%

Table~\ref{sdtab} lists all suspected subdwarfs (according to our original 
estimates approximately $v_\mathrm{t}>250$~km/s) with their roughly
corrected (reduced by 50\%) distances and tangential velocities.
The spectra of all these objects are shown in Fig.~\ref{11sd}. Note that 
five of them have trigonometric distances which correspond to 
30\% to 60\% of our original estimates (see Table~\ref{20obj} and 
Fig.~\ref{dvsd}). LHS~2281 is one more object in our sample 
with $v_\mathrm{t}>200$~km/s according to our original distance estimate
of 69~pc. This is, however, in good agreement with its Hipparcos distance
of 54~pc. The next smaller tangential velocity is measured for LHS~105
with about 170~km/s according to our original distance estimate of about
32~pc. Therefore, a threshold of $v_\mathrm{t}>250$~km/s seems to be 
justified for suspecting objects in our sample as subdwarfs.

\begin{table}
 \footnotesize
 \caption[]{Suspected subdwarfs with corrected distances
and tangential velocities
}
\label{sdtab}
 \begin{tabular}{lcrc}
 \hline
 \hline
Name & SpT$^\mathrm{corr}$ &$d^\mathrm{corr}$ & $v_\mathrm{t}^\mathrm{corr}$ \\
     &                     & [pc]             & [km/s] \\
 \hline
LP~552-21  & sd: M0.5 & 348 & 567 \\
LHS~5343   & sd: K7.0 & 187 & 444 \\
LHS~299    & sd: M0.0 &  75 & 397 \\
LP~323-168 & sd: K2.0 & 398 & 394 \\
LHS~2763   & sd: M0.0 &  82 & 309 \\
LHS~6124   & sd: K7.0 & 121 & 252 \\
LHS~6147   & sd: K7.0 & 116 & 231 \\
LHS~6385   & sd: K5.0 &  61 & 136 \\
LHS~2048   & sd: M2.5 &  52 & 131 \\
LHS~6356   & sd: K5.0 &  63 & 127 \\
LHS~3473   & sd: M3.0 &  33 & 124 \\
 \hline
 \end{tabular}
\end{table}

\subsection{White dwarfs}

\subsubsection{LHS~1200}
\label{wddist1}

A first rough distance estimate for the early-type DA white dwarf LHS~1200 is
based on our assumption that it has a similar absolute magnitude as the
two relatively hot white dwarfs Wolf~1346 and GD~140, both classified as
DA3 (Greenstein \cite{greenstein84}). 
These
two objects have
very similar Hipparcos (ESA \cite{esa97}) parallax measurements of about
68~mas and 65~mas for Wolf~1346 and GD~140, respectively.  
We compared the SSS $B_\mathrm{J}$ and $R$ magnitudes of Wolf~1346 and GD~140
($B_\mathrm{J}=11.90, R=12.03$ and $B_\mathrm{J}=12.02, R=12.09$) with those 
of LHS~1200 (see \S\ref{wdspec}). The resulting distance of LHS~1200 is
$d_{B_\mathrm{J}}=133$~pc or $d_R=148$~pc. Note the good agreement of these
two estimates, which is also due to the fact that all three objects have
nearly the same colour ($B_\mathrm{J}-R\sim-0.1$).
The mean distance of about 140~pc 
combined with the proper motion of about 0.55~arcsec/yr
yields a very high tangential velocity of about 370~km/s.
Such a velocity would classify LHS~1200 as a member of the Galactic halo.
Compared to the well known 
 halo DA7 white dwarf LHS~56 (Liebert, Dahn \& Monet \cite{liebert88}) 
at a distance of about 18~pc,
LHS~1200 is roughly 3~mag fainter in the optical (SSS) and exhibits an
about 7 times smaller total proper motion, in good agreement with the
expectations for a halo object being 8 times more distant.

A more accurate distance can be obtained from the Sloan Digital
Sky Survey (SDSS) photometry of LHS~1200. In the 3rd data 
release, DR3\footnote{http://www.sdss.org/dr3/}, we find $u=16.717, g=16.541,
r=16.872, i=17.119, z=17.388$. The SDSS colours agree well with those of 
DA3 and DA3.5 white dwarfs in Harris et al.\ (\cite{harris03}) and 
Kleinman et al.\ (\cite{kleinman04}). 
Using the photometric transformation of Fukugita et al.\ (\cite{fukugita96})
we obtain $V=16.715, V-R_\mathrm{c}=-0.032$ and $V-I_\mathrm{c}=-0.077$,
consistent with a $T_{\rm eff}=13000$~K, i.e. DA3.5 white dwarf with an
absolute magnitude of $M_\mathrm{V}\sim11.5$ (cf. Tab.~3 and Fig.~6
in Bergeron, Wesemael \& Beauchamp \cite{bergeron95}). Adopting the spectral
type of DA3.5 and assuming an uncertainty of 0.35~mag for the above given
absolute magnitude we compute a distance of $110\pm20$~pc. The corresponding
tangential velocity is about 290~km/s. Note that the alternative spectral type
of DA3 would allow a direct distance estimate using SDSS photometry of
three DA3 white dwarfs with measured Hipparcos parallaxes: Feige~22, G~93-48,
and Wolf~1346 (Smith et al.\ (\cite{smith02}; ESA \cite{esa97}). 
As expected, the resulting distance would be larger, $\sim$150~pc, 
and the tangential velocity still extremely large.

The Tautenburg spectrum of LHS1200, covering a smaller wavelength interval
but having a better spectral resolution than the CAFOS spectrum shown in 
Fig.~\ref{2wd}, was used for a rough estimate of the radial velocity by 
measuring the wavelengths of the line cores of the Balmer lines H$\beta$ 
and H$\gamma$ relative to sky background lines in the same wavelength range. 
For both lines the same uncorrected shift of +2.5 ($\pm 1$)\AA\, is measured. 
If we apply the required gravitational redshift correction,
which is of the order of $-30$~km/s (Silvestri et al.\ \cite{silvestri01}),
we derive a heliocentric radial velocity of about $+115 \pm 60$~km/s.
Using the latter and the distance of $110\pm20$~pc, we compute a heliocentric 
space velocity following Johnson \& Soderblom~(\cite{johnson87}) 
of ($U, V, W = -116\pm32, -184\pm63, -219\pm54$)~km/s.
Despite the large errors, this very high space velocity
supports a Galactic halo membership of LHS~1200. In particular,
it does not share the Galactic rotation speed of the local spiral
arm ($V$), and it also shows a large velocity perpendicular to the
Galactic plane ($W$).

\subsubsection{LHS~2288}
\label{wddist2}

For LHS~2288, which we classified as a cool white dwarf, we first apply the 
photographic absolute magnitude-colour relation
\begin{equation}
\label{photd}
M_{B_\mathrm{J}} = 12.73 + 2.58(B_\mathrm{J}-R),
\end{equation}
derived by Oppenheimer et al.\ (\cite{oppenheimer01}) and obtain
a photometric distance of 41~pc. 
Again we are lucky enough to find this object in the SDSS so that we can
derive a more accurate photometric distance. The SDSS magnitudes, $u=22.259,
g=20.065, r=18.921, i=18.541, z=18.325$ can be transformed to $V=19.454,
V-R_\mathrm{c}=+0.790, V-I_\mathrm{c}=+1.412$
(Fukugita et al.\ \cite{fukugita96}).
Using the $M_\mathrm{V}$ vs. $V-I_\mathrm{c}$ relationship for cool white
dwarfs from Salim et al.\ (\cite{salim04}; their Equation 7), we obtain
an absolute magnitude of $M_\mathrm{V}=16.23$. Assuming again an uncertainty
of 0.35~mag in the absolute magnitude, we get a distance of 44$\pm$7~pc,
in very good agreement with the previous distance estimate.

Combined with the proper motion of about
0.68~arcsec/yr, this yields a more moderate tangential velocity 
of about 140~km/s for LHS~2288, 
probably placing this object in the Galactic thick disk
population. Compared to the sample presented in Oppenheimer 
et al.\ (\cite{oppenheimer01}), this object seems to be a relatively 
nearby representative of the cool white dwarfs.
From its SDSS colours it resembles SDSS~J0854$+$35, an ultracool white 
dwarf with only mild collision-induced absorption 
(Gates et al.\ \cite{gates04}), with LHS~2288 being located at 
slightly shorter distance.

\section{Notes on individual objects}
\label{indobj}

Included in Table~\ref{20obj} are flags, indicating that additional notes
are available for these stars. These notes are collected in a separate
file supplementing the corresponding data file of Table~\ref{20obj},
available via the CDS. The notes are mainly dealing with spectral 
features mentioned during the inspection of the low-resolution spectra, 
e.g. (weak indication of) H$\alpha$ emission lines (43 stars), but
do also include brief information on common proper motion companions.

In the following subsections we consider some of the most remarkable
and interesting (groups of) individual stars in our sample in more
detail.

\subsection{
Sample
stars within 10 pc}
\label{10pcstars}

According to our spectroscopic distances there are in total eight stars 
within 10~pc, usually in good agreement with independent determinations.
Only three of these objects had previously announced spectroscopic 
distances. However, all others were known as very nearby stars from recent
photometric distance determinations.
 
The nearest one, the M6.5 dwarf LHS~2090, was already published in paper~I 
with a first spectroscopic distance estimate of 6.0$\pm$1.1~pc. Reid \&
Cruz (\cite{reid02a}) determined a photometric distance of 5.2~pc.
It's preliminary trigonometric parallax of $159.02 \pm4.46$~mas
on the RECONS homepage\footnote{http://www.chara.gsu.edu/RECONS} 
confirms very nicely our new spectroscopic distance of 6.3~pc.
 
Available optical photometry yields a distance of 7.1~pc 
(Jahrei{\ss} \cite{jahreiss05}) for the second nearest of our sample, the 
M5 dwarf G~161-71, very close to the 6.9~pc given in Table~\ref{20obj}.
Note that a spectral type of M5Ve was already assigned to this star
by Torres (\cite{torres00}).
 
LHS~6167 (M5 at 7.3~pc) is most probably identical with the faint X-ray 
source 1RXS~J091535.4-103538 (Voges et al.\ \cite{voges00}).
Reid et al. \cite{reid02b} found a similar photometric distance of 
6.7~pc. A spectral type M4-5 was previously assigned to this star
by Gigoyan et al.\ (\cite{gigoyan98}).
 
For LP~775-31 (M7.5 at 8.2~pc) there are several independent distance 
determinations in the literature. Reid \& Cruz (\cite{reid02a}) obtained
a photometric distance of 6.4~pc. whereas  
Cruz \& Reid (\cite{cruz02}) initially typed this star as M6 at 11.3~pc 
but later revised it to M7 at 8.6~pc in Cruz et al. (\cite{cruz03}).
In paper~II, a much later spectral type of M8 and a correspondingly 
shorter distance of 6.4~pc were derived. Reid (\cite{reid03c}) also lists 
this star with a spectral type of M8.
 
For LP~870-65 (M4.5 at 8.7~pc) available optical photometry yields a slightly 
larger distance of 10.7~pc (Jahrei{\ss} \cite{jahreiss05}), whereas 
Reid et al.\ (\cite{reid03b}) found a photometric distance of 10.1~pc.
 
For LP~876-10 (M4 at 8.7~pc) available optical photometry 
yields 7.8~pc (Jahrei{\ss} \cite{jahreiss05}), and 
Reid et al.\ (\cite{reid03b}) found even 7.2~pc as
photometric distance.
 
For LHS~1901 (M6 at 9.3~pc) Reid et al.\ (\cite{reid03b}) found 
a photometric distance of 7.8~pc and a spectroscopic distance of
8.0~pc by classifying this star as M6.5 dwarf.
 
LP~869-26 (M5 at 9.4~pc) may be identical with the bright X-ray source 
1RXS J194453.7-233743 (Voges et al.\ \cite{voges99}). Optical photometry 
yields 10.7~pc in Jahrei{\ss} (\cite{jahreiss05}), whereas 
Reid et al.\ (\cite{reid03b}) found a similar photometric distance of 9.2~pc.

One object, LP~869-19, with a previous photometric distance of 9.8~pc as 
determined by Reid et al.\ (\cite{reid03b}) is according to our spectral 
classification as an M4 dwarf placed slightly further away, at 11.6~pc.

\subsection{Binaries and common proper motion stars}
\label{cpms}

About 6\% of the stars in our sample are apparently components of
(wide) binaries, which usually appear in Luyten's catalogue of double 
stars with common proper motion (LDS; Luyten \cite{luyten4087}).
All these stars are also marked by $^*$ in Table~\ref{20obj} and have
a brief comment in the separate notes file supplementing the
electronic table available at the CDS:

G158-073 (M2.5 at 23~pc) and G158-074 (M4 at 30~pc), LDS~3163 
(sep. 15~arcsec). The independent spectroscopic distance estimates
of the two components are in reasonable good agreement.
G158-073 has a photometric distance of 26~pc in 
Reid et al.\ (\cite{reid03b}).

CD$-$25$^{\circ}$1159 (M2 at 17~pc) is a close binary of nearly
equally bright components detected by Hipparcos
(sep. 0.3~arcsec; ESA \cite{esa97}). We did not resolve the binary
in our spectroscopic observations. Therefore, a correcting factor of 1.4 has 
to be applied to our distance estimate, which is then in better agreement
with the Hipparcos distance to the system (28~pc).

LP~831-14 (M2.5 at 28~pc) and the bright F2V star Zeta For
 (not part of our survey, with a Hipparcos parallax of 
30.92$\pm$0.77~mas; ESA \cite{esa97}) form 
the pair LDS~3446 (sep. 176~arcsec).
Their proper motions are slightly different 
(Salim \& Gould \cite{salim03}). Our spectroscopic distance
estimate agrees well with the Hipparcos measured distance.

G221-027 (M4 at 17~pc; classified as M4 at 15~pc by  Cruz \& 
Reid \cite{cruz02}) and LP~31-302 (not part of our survey, but
classified by Cruz \& Reid (\cite{cruz02}) as M5 at 13~pc), 
LDS~1589 (sep. 5~arcsec). The various distance estimates 
are in good agreement with each other.

LP~359-216 (M2.5 at 12~pc) and its fainter companion LP~359-186
(not part of our survey), LDS~6160 (sep. 167~arcsec).
LP~359-216 was originally selected by us as a nearby candidate
according to its Tycho-2 proper motion and photometry compared
to its 2MASS magnitudes. It is the nearest M2.5 dwarf in our sample.
We identify LP~359-216 
with the bright 
X-ray source 1RXS~J050315.1$+$212351 (Voges et al.\ \cite{voges99})
and HD~285190. Simbad, however, gives a spectral type of
K0 for the latter 
and does not associate the two names with each other.
Our spectrum of LP~359-216 (on top of Fig.~\ref{11early}) clearly
rules out a K-type object.
The trigonometric parallax of 36.5$\pm$8.6~mas for
LP~359-216 (van Altena, Lee \& Hoffleit \cite{vanaltena95})
is in strong contrast with our much closer distance estimate,
even if we take into account the relatively large error of
the trigonometric parallax. If we assume LP~359-216 and LP~359-186 
be associated and prefer our spectroscopic distance estimate to 
the trigonometric one, 
the projected physical separation of the system
reduces from 
$\sim$4600~AU to $\sim$2000~AU. The 2MASS colours and magnitudes 
of LP~359-186 ($J-K_\mathrm{s}=+0.86, J=9.75$) are consistent with 
an M5 dwarf secondary at 12~pc distance.
In view of these facts the trigonometric parallax seems to be debatable.

LP~784-12 (M1.5 at 21~pc) is a close binary measured by Hipparcos 
(sep. 0.37~arcsec; ESA \cite{esa97}). We did not resolve the binary
in our spectroscopic observations. Therefore, a correcting factor of up to
1.4 has to be applied to our distance estimate, which is then in better 
agreement with the Hipparcos distance to the system (29~pc).

LP~729-54 (M4 at 13~pc) and the fainter companion LP~729-55 
(not part of our survey),  LDS~3977 (sep. 14.5~arcsec).
Reid et al.\ (\cite{reid03b}) give a spectral type of M3.5
with a distance of 50~pc for LP~729-55.  We suppose that
they have assigned the spectral type to the wrong object, yielding
a much too large distance to the system.

LP~790-47 (K5 at 62~pc) and its fainter companion LP~790-46
(not part of our survey), LDS~3984 (sep. 24.5~arcsec).
Both stars have accurate proper motion measurements in the
UCAC2 (Zacharias et al. \cite{zacharias04}), which agree 
within the errors. The pair appeared in the study of
halo common proper motion stars by Ryan (\cite{ryan92}), but
was not further investigated, since red stars were excluded
from the analysis. Our moderately large value of the tangential
velocity determined for LP~790-47 ($\sim$60~km/s) hints at a
thick disk membership of the pair.

BD$+$26$^{\circ}$2161A (K2 at 23~pc) and its slightly fainter companion
BD$+$26$^{\circ}$2161B (also part of our survey but not observed 
spectroscopically)
are mentioned
in the NLTT
notes (sep. 5~arcsec). The spectrum of BD$+$26$^{\circ}$2161A is shown 
in Fig.~\ref{11early}. This pair could already have been suspected as 
nearby from Hipparcos observations, since both stars have positive 
Tycho parallaxes, 
which in case of the primary is highly significant: 63.1$\pm$11.1~mas,
whereas the value for the secondary is smaller: 33.1$\pm$17.1~mas
(ESA \cite{esa97}). The primary is not included in the list of
suspected nearby Tycho stars of Makarov (\cite{makarov97}), although
it seems to fulfil his selection criteria.

LP~852-5 (M3 at 24~pc) and its 2.2~mag fainter companion LP~852-6
(not part of our survey), LDS~4178 (sep. 14~arcsec).
The 2MASS colours and magnitudes of the secondary ($J-K_\mathrm{s}\sim+0.86,
J=11.47$) are consistent with a mid-M spectral type ($\sim$M5.5)
at $\sim$25~pc.

LP~65-440 (M4 at 19~pc) is one component of the unresolved 
(also in 2MASS) binary LDS~2674 (sep. 1.5~arcsec). The distance
may therefore be up to 1.4 times larger (in case of an equal
spectral type of both components).

LP~797-105 (M2.5 at 33~pc) has a bright companion, LP~797-104
(not part of our survey, sep. 138~arcsec) with a Hipparcos parallax of 
27.4$\pm$1.6~mas, in good agreement with our spectroscopic distance
of LP~797-105. The UCAC2 proper motions of the two objects almost
agree within their errors (Zacharias et al.\cite{zacharias04}).

LP~439-442 (M3 at 24~pc) and its 1.2~mag fainter companion LP~439-441
(not part of our survey), LDS~1415 (sep. 15~arcsec).
The much shorter photometric distance estimate (13~pc) for LP~439-442
by Reid \& Cruz (\cite{reid02a}) probably results from an error 
in reading the correct $V$ magnitude from Ryan (\cite{ryan92}, i.e.
$V=13.1$, and not $V=13.95$ as used by Reid \& Cruz.
The 2MASS colours and magnitudes of the secondary ($J-K_\mathrm{s}\sim+0.87,
J=10.46$) are fully consistent with an $\sim$M4 dwarf at $\sim$25~pc.

LP~806-5 (M2.5 at 22~pc) has a slightly (0.1~mag) brighter wide 
companion LP~806-4 (not part of our survey, sep. 197~arcsec),
which could have the same spectral type. Both components have
accurate proper motion measurements in the UCAC2 
(Zacharias et al.\ \cite{zacharias04}), which agree within their
errors.

LHS~3473 (M3 at 66~pc) and its only slightly (0.3~mag) fainter
companion LHS~3474 (sep. 4.5~arcsec) are unresolved on
Schmidt plates. A spectrum was observed for LHS~3473 only.
The spectrum, taken at a seeing of about 1 arcsec, 
is clearly unaffected by the close companion.
The trigonometric parallax of LHS~3473 ($33.3\pm3.0$~mas)
listed by van Altena, Lee \& Hoffleit (\cite{vanaltena95})
places the system at only 30~pc distance. 
Therefore, we assume LHS~3473 to be a subdwarf.
The two objects are well resolved in 2MASS and have similar NIR colours.
Therefore, the system
probably consists of two subdwarfs with a projected physical separation 
of 135~AU and a tangential velocity of $\sim$110~km/s with respect to the Sun.

LHS~6356 (K5 at 126~pc) has a trigonometric distance ($d=68$~pc;
van Altena, Lee \& Hoffleit \cite{vanaltena95})
only half as large as our spectroscopic one, indicating that it is in fact
a subdwarf. Its common proper motion companion, LHS~6357 (sep. 8~arcsec
according to the NLTT notes), is still lacking a spectral type, but
has a similar although uncertain trigonometric parallax of $20\pm5$~mas
(Harrington \& Dahn \cite{harrington80}). LHS~6357 is roughly 0.8~mag
fainter in the NIR (2MASS). It may also be a subdwarf, but of a
later spectral type. 

\subsection{Dubious common proper motion pairs}
\label{wrongcpm}

LP~932-83 (M5 at 10~pc) and its brighter companion Hip~112648
(not part of our survey, Hipparcos parallax: 23$\pm$2~mas), 
LDS~4999 (sep. 217~arcsec). The Hipparcos distance of the 
K7 primary is four times larger than our spectroscopic distance
of the secondary.  The physical association of the two objects may be not 
real, although their proper motions almost agree within the errors 
(Zacharias et al.\ \cite{zacharias04}, Salim \& 
Gould \cite{salim03}). We note that Reid \& Cruz (\cite{reid02a})
determined a photometric distance of 14~pc for LP~932-83, in good 
agreement with our spectroscopic distance estimate. 
Even if we assume LP~932-83 to be a close binary a discrepancy
with the distance of Hip~112648 remains.

The object with our largest estimated spectroscopic distance ($\sim$800~pc),
the K2 star LP~323-168, 
which we suspect to be a subdwarf with a distance more like $\sim$400~pc,
has an apparent common proper 
motion companion, LP~323-169, at an angular separation of about 3~arcmin,
as claimed in the NLTT remarks. In strong contrast to the assumed physical
association of the two objects, we classify the latter as 
a nearby ($d\sim14$~pc) M4.5 dwarf. According to our own 
determination (using positions from the APM catalogue of McMahon, Irwin \& 
Maddox \cite{mcmahon00}, the SSS and the 2MASS) and according to the
USNO CCD Astrograph Catalog (UCAC2; Zacharias et al.\ \cite{zacharias04})
the two objects have clearly different large proper motions. It is just
by chance that two different high proper motion stars, 
a distant high-velocity representative of the Galactic halo
and a member of the Solar neighbourhood,
appear in the same small sky field. 

LP~731-76 (M5 at 11~pc) and its bright apparent companion 
BD$-$10$^{\circ}$3166
(not part of our survey), LDS~4041 (sep. 17~arcsec). Although
the K0 primary has a long bibliography in the Simbad data base, where
it is also mentioned as a planetary host star 
(Butler et al.\ \cite{butler00}), the star (system) did not yet get 
attention for its proximity. The common proper motion is doubtful, because
the UCAC2 proper motions (Zacharias et al.\ \cite{zacharias04}) 
of the two stars are significantly different, in particular in
declination ($-93$ vs. $-6$~mas/yr). Finally, a physical association
of the two objects can be ruled out on the basis of a photometric
distance estimate for BD$-$10$^{\circ}$3166. 
Using $V=10.08, B-V=+0.84, U-B=+0.58$
and the optical absolute magnitude-colour relation described in the  
forthcoming CNS4 (Jahrei{\ss} \cite{jahreiss05}),
we derive a photometric distance of about 64~pc for this K0V star,
six times larger than our spectroscopic distance to LP~731-76. 

LP~709-85 (M3 at 26~pc) and its fainter companion LP~709-68
(not part of our survey), LDS~3366 (sep. 150~arcsec),
have significantly different proper motions (Zacharias
et al.\ \cite{zacharias04}), which may question their physical
relationship at the given distance.

\subsection{
A possible new pair that was rejected}
\label{rejectcpm}

During the process of cross-identification of our sample stars with those
of other recent distance determinations we realised that LP~931-53
(M2.5 at 26~pc) seems to have a common proper motion companion,
LP~931-54, which was found by Reid et al.\ (\cite{reid03b}) to lie at
approximately the same distance (32~pc). The angular separation between
the two NLTT stars is about 15~arcmin, their proper motions 
$(\mu_{\alpha}\cos{\delta},\mu_{\delta})$ as given in
the NLTT are $(-289,-333)$ and $(-133,-386)$~mas/yr for LP~931-53 and 
LP~931-54, respectively. Much more similar values, 
$(-286,-342)$ and $(-285,-387)$~mas/yr, can be found in the Yale/San Juan 
Southern Proper Motion catalogue (Platais et al.\ \cite{platais98}).
We have checked the proper motions of the two stars using all available
SSS data (8 different epochs) combined with 2MASS and DENIS 
(Epchtein et al.\ \cite{epchtein97}) positions, yielding
$(-286,-350)$ and $(-150,-411)$~mas/yr with errors smaller than 10~mas/yr.
Since these values confirm the different proper motions originally
reported in the NLTT, we therefore reject our initial assumption that
the two objects form a wide binary at the given distance and
angular separation. However, they may be members of a stellar stream
in the Solar neighbourhood.

\subsection{Nearby candidates from Fleming (\cite{fleming98})}
\label{xflem}

As already mentioned in \S~\ref{addsel}, we have included a few stars
previously found as nearby candidates by Fleming (\cite{fleming98})
based on their X-ray fluxes. 
The above mentioned M4.5 star, LP~323-169, as well as two other
objects  (G179-055; M4 at 14~pc and G224-065; M3.5 at 26~pc) 
were originally selected from the list of nearby candidates found by 
Fleming (\cite{fleming98}). One more object contained in Fleming's
list was selected independently by us as a nearby candidate and confirmed
as such by our spectroscopy (LP 463-23; M5.5 at 14~pc). In all four
cases our spectroscopic distance estimates place these stars within or
close to the 25~pc radius, in good agreement with the photometric distances 
predicted by Fleming. G179-055 was found to be a spectroscopic binary
(Mochnacki et al.\ \cite{mochnacki02}). 
Further stars from Fleming (\cite{fleming98}),
with proper motions below the NLTT proper motion limit of 0.18~arcsec/yr, 
will be described in paper IV.

\subsection{
Three 
nearby halo stars?}
\label{nearhalo}

There are two stars, LHS~269 and LHS~436, which have spectroscopic distances 
within 25~pc and tangential velocities larger than 100~km/s. Both 
have been classified as mid-M dwarfs (M5 and M5.5). A very large tangential 
velocity ($\sim175$~km/s) was obtained for the M4 dwarf LHS~105 at about 
32~pc distance. Such high velocities may indicate a Galactic halo membership
of these normal M dwarfs, although radial velocities are still lacking,
and the space velocities could also be consistent with the thick disk
population. Other normal (non-metal-poor) mid-M dwarfs with very large
space motions were recently reported by  L\'epine, Rich \& 
Shara (\cite{lepine03}).

\subsection{Nearby K dwarfs or distant 
(sub)giants?
}
\label{KVorIII}

The earliest-type (K1) object in our sample of NLTT stars, 
BD$+$32$^{\circ}$588 (Fig.~\ref{11early}), was 
initially
identified as
a nearby ($d\sim22$~pc) dwarf. The Hipparcos parallax, corresponding to a
much larger distance ($d\sim190$~pc), of this rather bright ($J\sim6.2$) star 
only allows the alternative interpretation that it is in fact a high-velocity
($v_\mathrm{t}\sim170$~km/s) early-K 
giant (see also \S\ref{KMdist}). 
For other K stars, classified by us as
nearby objects, our distances agree well with previous trigonometric
parallax measurements. However, there are two other bright early-K
stars, BD$+$26$^{\circ}$2161A ($J\sim6.4$) and G204-045 ($J\sim6.8$), 
placed by our spectroscopic distances within 25~pc. Their spectra are 
also shown in Fig.~\ref{11early}. These two stars are still lacking accurate
trigonometric parallaxes 
(the Tycho parallaxes of 
BD$+$26$^{\circ}$2161A and its common proper motion companion
have errors of $\sim$20\% and $\sim$50\%, respectively; 
see \S\ref{cpms}). We can not exclude their possible 
(sub)giant nature.
All other objects in our sample are fainter and of later spectral type. 
Therefore, we do not expect to find more (sub)giants among these high 
proper motion stars.

\section{Conclusions}
\label{concdisc}

Compared to the expected large numbers of missing stellar systems
within the 25~pc horizon - 
about 2000 stars according to Henry et al.\ (\cite{henry02}) - 
our study has investigated and uncovered 
only a small fraction of the unknown close neighbours to the Sun. 
Clearly, further efforts are needed to complete the 25~pc sample.

Among the nearest objects found, there are eight stars with
distances smaller than 10~pc (see  \S\ref{10pcstars}). For five of them, 
spectroscopic distances are published here for the first time. 
There are eight objects without any previous distance estimates 
that we find to be within 15 pc: 
LP~ 780-32 (M4 at 10.7~pc), LP~64-185 (M4 at 12.3~pc), LP~331-57
(M3 at 12.7~pc), LHS~5140 (M4 at 13.0~pc), G100-050 (M5 at 13.6~pc),
LP~278-42 (M5.5 at 13.8~pc), G123-045 (M4 at 14.1~pc), and 
G036-026 (M5 at 14.3~pc). We have to keep in mind that some of our M3 
and M4 dwarfs may be in fact M3.5 dwarfs (see \S\ref{clsMK}) and therefore
located slightly nearer or farther, respectively.
Out of all 322 objects in our sample,
138 did not have any previous distance determination.

With few exceptions, mainly concerning suspected sub\-dwarfs,
our distance estimates agree well with previously
determined values. The distances of components in common proper motion
systems are generally consistent with each other. However, there are two 
cases, LDS~6160 and LDS~4999, which need an additional check of the results,
since our spectroscopic distance disagrees with a trigonometric distance.
The distant K2 star LP~323-168 and the nearby M4.5 dwarf LP~323-169, mentioned
in the NLTT remarks as a possible common proper motion pair, are
definitely not physically associated. LP~731-76 and BD$-$10$^{\circ}$3166
(LDS~4041) can almost certainly be ruled out as a physical pair.

The maximum tangential velocity left after a 50\% reduction of the
largest distances in our sample is about 570~km/s, i.e. almost within the
usual range for subdwarfs. The tangential velocities of the
K and M subdwarfs listed in Gizis (\cite{gizis97}) are all smaller than
500~km/s (typically between 200 and 400~km/s). An extremely large tangential
velocity (580~km/s) was determined for the recently discovered sdM1.5
object, SSSPM~J1530$-$8146 (Scholz et al.\ \cite{scholz04}).
There is, however, one object
with an uncertain sdK7 spectral type, at the spectroscopic distance of about
375~pc and with a tangential velocity larger than 1300~km/s reported by
L\'epine, Rich \& Shara (\cite{lepine03}). With such an extra-ordinary high
velocity this latter object, LSR~0400$+$5417, belongs no longer to
the Galaxy.
Adopting here again an absolute magnitude of $M_\mathrm{J}=6.9$ for sdK7 stars, 
we
compute a somewhat smaller distance of 274~pc for this object based on
its 2MASS $J$ magnitude,
which was not
considered by L\'epine, Rich \& Shara (\cite{lepine03}). This reduces the
tangential
velocity of LSR~0400$+$5417 to about 950~km/s, which is still a record value.

The relatively hot ($\sim$DA3.5) white dwarf LHS~1200 is so faint
that it must be located farther than 100~pc. A conservative distance estimate
of 110$\pm$20~pc based on SDSS photometry, its large proper motion
and our radial velocity measurement lead to a very high heliocentric
space velocity typical of Galactic halo members. We can not exclude
that it is even hotter (DA3), and consequently at an even larger
distance of about 150~pc, and with an even more extreme space velocity.  
Further investigation of this interesting object 
is
needed.

About 85\% of our M and K stars are found to lie within 30~pc, and 72\% are
within 25~pc. This is a
high success rate, if we consider our preliminary photometric distances
used for the selection of the spectroscopic targets as rough estimates only.
We must also take into
account that about one third of the objects beyond 30~pc were selected
for other reasons than for their optical-to-infrared colours and for
their corresponding photometric distance estimates.
The high success rate indicates that a purely photometric selection based on
all available photographic optical magnitudes combined with the meanwhile
completed 2MASS data base may be worth doing. 

\begin{acknowledgements}

The spectroscopic confirmation of nearby star candidates is mainly based on
observations made with the
2.2\,m telescope of the German-Spanish Astronomical Centre, Calar Alto, Spain.
Part of the observations were carried out in service mode. We would
like to thank the Calar Alto staff for their 
kind support and for their help with the observations.
The spectrum of one of the new 10pc stars was taken with the Italian
Telescopio Nazionale Galileo (TNG) operated at the island of La Palma by the
Centro Galileo Galilei of the INAF (Instituto Nazionale die Astrofisica)
at the Spanish Observatorio del Roque de los Muchachos of the Instituto de
Astrofisica de Canarias.
The Tautenburg spectrum of LHS~1200 was kindly provided by Bringfried Stecklum.
HM acknowledges financial support from the Deutsche Forschungsgemeinschaft
for numerous observing campaigns at Calar Alto and from the European
Optical Infrared Coordination Network for Astronomy, OPTICON.

We acknowledge the use of images from the Digitized Sky Survey, produced at 
the Space Telescope Science Institute under U.S. Government grant NAG W-2166. 
The images of these surveys are based on photographic data obtained using the 
Oschin Schmidt Telescope on Palomar Mountain and the UK Schmidt Telescope. 
We would also like to acknowledge the use of the batch interface to DSS
available at the Astronomical Data Analysis Center of the 
National Astronomical Observatory of Japan 
(http://dss.mtk.nao.ac.jp/).

This research has made use of data products from the SuperCOSMOS Sky Surveys
at the Wide-Field Astronomy Unit of the Institute for Astronomy, University
of Edinburgh. We have also used
data products from the Two Micron All Sky Survey, which is a joint project of
the University of Massachusetts and the Infrared Processing and Analysis
Center/California Institute of Technology, funded by the National Aeronautics
and Space Administration and the National Science Foundation.
We acknowledge the use of the Simbad data base and the VizieR Catalogue Service
operated at the CDS.

Artie Hatzes is acknowledged for helpful comments. Finally,
we would like to thank Darja Golikowa and Kai Schmitz for their help with 
the re-identification of high proper motion stars in DSS and SSS images,
the cross-correlation of USNO A2.0, SSS, and 2MASS data, and with
first distance estimates based on the photometry of
stars with measured trigonometric parallaxes.
We would also like to thank the referee, Kelle Cruz, for her very
detailed report with many important suggestions that lead to
improvements of the paper.
\end{acknowledgements}


%

%

\end{document}